%% file: hoecker_tau2010_arxiv.tex
\documentclass[nofootinbib,twocolumn,aps,superscriptaddress,preprintnumbers]{myRevtex4}
\pdfoutput=1

\usepackage{amsmath,amssymb,amsfonts} 
\usepackage{graphicx}
\usepackage{color}
\usepackage{xspace}
\usepackage{natbib}   
\usepackage{tabularx}
\usepackage{hyperref}

\definecolor{darkblue1}{rgb}{0,0,.4}
\hypersetup{breaklinks=true, 
            colorlinks=true, 
            linkcolor=darkblue1, 
            menucolor=darkblue1, 
            urlcolor=darkblue1,
            citecolor=darkblue1
}

\bibstyle{plain}
\input defs
\input journals

\newcommand{\figsize}{\columnwidth}
\newcommand{\fighspace}{0.4cm}

\begin{document}

\preprint{\vbox{\hbox{CERN-OPEN-2010-026}}}

\title{The Hadronic Contribution to the Muon Anomalous Magnetic Moment \\
       and to the Running Electromagnetic Fine Structure Constant at \boldmath$M_Z$ --- \\
       Overview and Latest Results}

\author{Andreas Hoecker\\[0.1cm]
     {\em\small CERN, CH--1211, Geneva 23, Switzerland}}

\date{\today}

\input abstract

\maketitle

\input bulk
\end{document}

%% file: defs.tex
\newcommand{\cbar}  {\ensuremath{\overline c}\xspace}

\mathchardef\Upsilon="7107
\def\Y#1S{\ensuremath{\Upsilon{(#1S)}}\xspace}

\newcommand{\mZ}{\ensuremath{M_Z}\xspace}
\newcommand{\as}{\ensuremath{\alpha_{\scriptscriptstyle S}}\xspace}

\newcommand{\ass}{\ensuremath{\as(s)}\xspace}
\newcommand{\asZ}{\ensuremath{\as(\mZ^2)}\xspace}

\newcommand{\aZ}{\ensuremath{\alpha(M_Z^2)}\xspace}
\newcommand{\dahadZ}{\ensuremath{\Delta\alpha_{\rm had}(M_Z^2)}\xspace}
\newcommand{\dahadZf}{\ensuremath{\Delta\alpha_{\rm had}^{(5)}(M_Z^2)}\xspace}

\newcommand{\Kbar    }{\kern 0.2em\overline{\kern -0.2em K}{}\xspace}
\newcommand{\Dbar    }{\kern 0.2em\overline{\kern -0.2em D}{}\xspace}

\newcommand{\Kz      }{\ensuremath{K^0}\xspace}
\newcommand{\Kzb     }{\ensuremath{\Kbar^0}\xspace}
\newcommand{\KzKzb   }{\ensuremath{\Kz \kern -0.16em \Kzb}\xspace}
\newcommand{\Kp      }{\ensuremath{K^+}\xspace}
\newcommand{\Km      }{\ensuremath{K^-}\xspace}

\newcommand{\KpKm    }{\ensuremath{\Kp \kern -0.16em \Km}\xspace}
\newcommand{\KS      }{\ensuremath{K^0_{\scriptscriptstyle S}}\xspace} 
\newcommand{\KL      }{\ensuremath{K^0_{\scriptscriptstyle L}}\xspace}

\def\CPT               {\ensuremath{C\!PT}\xspace} 

\newcommand{\mc}{\multicolumn}

\newcommand{\Tau}{\ensuremath{\tau}\xspace}
\newcommand{\nut}{\ensuremath{\nu_\tau}\xspace}

\newcommand{\BR}{\ensuremath{{\cal B}}\xspace}

\newcommand{\pip}{\ensuremath{\pi^+}\xspace}
\newcommand{\pim}{\ensuremath{\pi^-}\xspace}
\newcommand{\piz}{\ensuremath{\pi^0}\xspace}

\newcommand{\ee}{\ensuremath{e^+e^-}\xspace}

\newcommand{\pp}{\ensuremath{\pi^+\pi^-}\xspace}
\newcommand{\pipiz}{\ensuremath{\pi^-\pi^0}\xspace}

\newcommand{\tev}{\ensuremath{\mathrm{\,Te\kern -0.1em V}}\xspace}
\newcommand{\gev}{\ensuremath{\mathrm{\,Ge\kern -0.1em V}}\xspace}
\newcommand{\mev}{\ensuremath{\mathrm{\,Me\kern -0.1em V}}\xspace}
\newcommand{\kev}{\ensuremath{\mathrm{\,ke\kern -0.1em V}}\xspace}
\newcommand{\ev}{\ensuremath{\mathrm{\,e\kern -0.1em V}}\xspace}
\newcommand{\gevc}{\ensuremath{{\mathrm{\,Ge\kern -0.1em V\!/}c}}\xspace}
\newcommand{\mevc}{\ensuremath{{\mathrm{\,Me\kern -0.1em V\!/}c}}\xspace}
\newcommand{\gevcc}{\ensuremath{{\mathrm{\,Ge\kern -0.1em V\!/}c^2}}\xspace}
\newcommand{\mevcc}{\ensuremath{{\mathrm{\,Me\kern -0.1em V\!/}c^2}}\xspace}

\newcommand{\bei}{\begin{itemize}}
\newcommand{\eei}{\end{itemize}}
\newcommand{\ben}{\begin{enumerate}}
\newcommand{\een}{\end{enumerate}}
\newcommand{\beq}{\begin{equation}}
\newcommand{\eeq}{\end{equation}}
\newcommand{\beqn}{\begin{eqnarray}}
\newcommand{\eeqn}{\end{eqnarray}}
\newcommand{\beqns}{\begin{eqnarray*}}
\newcommand{\eeqns}{\end{eqnarray*}}

\renewcommand{\arraystretch}{1.25}
\newcommand{\amu}{\ensuremath{a_\mu}\xspace}
\newcommand{\amuhad}{\ensuremath{\amu^{\rm had}}\xspace}
\newcommand{\amuhadLO}{\ensuremath{\amu^{\rm had,LO}}\xspace}

\newcommand{\amuSM}{\ensuremath{\amu^{\rm SM}}\xspace}

\newcommand{\ie}{{i.e.}\xspace}

\newcommand{\cf}{{cf.}\xspace}
\newcommand{\ea}{{\em et al.}\xspace}
\newcommand{\ibid}{{\em ibid.}\xspace}
\newcommand{\taum}{\tau^-}

\def\@citex[#1]#2{\if@filesw\immediate\write\@auxout{\string\citation{#2}}\fi
  \@tempcnta\z@\@tempcntb\m@ne\def\@citea{}\@cite{\@for\@citeb:=#2\do
    {\@ifundefined
       {b@\@citeb}{\@citeo\@tempcntb\m@ne\@citea
        \def\@citea{,\penalty\@m\ }{\bf ?}\@warning
       {Citation `\@citeb' on page \thepage \space undefined}}%
    {\setbox\z@\hbox{\global\@tempcntc0\csname b@\@citeb\endcsname\relax}%
     \ifnum\@tempcntc=\z@ \@citeo\@tempcntb\m@ne
       \@citea\def\@citea{,\penalty\@m}
       \hbox{\csname b@\@citeb\endcsname}%
     \else
      \advance\@tempcntb\@ne
      \ifnum\@tempcntb=\@tempcntc
      \else\advance\@tempcntb\m@ne\@citeo
      \@tempcnta\@tempcntc\@tempcntb\@tempcntc\fi\fi}}\@citeo}{#1}}

\def\@citeo{\ifnum\@tempcnta>\@tempcntb\else\@citea
  \def\@citea{,\penalty\@m}%
  \ifnum\@tempcnta=\@tempcntb\the\@tempcnta\else
   {\advance\@tempcnta\@ne\ifnum\@tempcnta=\@tempcntb \else
\def\@citea{--}\fi
    \advance\@tempcnta\m@ne\the\@tempcnta\@citea\the\@tempcntb}\fi\fi}
\newenvironment{myquote}
               {\list{}{\leftmargin0cm\indent}%
                \item\relax}
               {\endlist}
\newcommand\allFontSize{\footnotesize}
\newcommand\detailsSize{\allFontSize}
\newenvironment{details}%
{\begin{myquote}\detailsSize}{\end{myquote}}


%% file: journals.tex
\def\anp#1,#2(#3){{\rm Adv.\ Nucl.\ Phys.\ }{\rm #1}, {\rm#2} {\rm(#3)}}
\def\aip#1,#2(#3){{\rm Am.\ Inst.\ Phys.\ }{\rm #1}, {\rm#2} {\rm(#3)}}
\def\aj#1,#2(#3){{\rm Astrophys.\ J.\ }{\rm #1}, {\rm#2} {\rm(#3)}}
\def\ajs#1,#2(#3){{\rm Astrophys.\ J.\ Supp.\ }{\rm #1}, {\rm#2} {\rm(#3)}}
\def\ajl#1,#2(#3){{\rm Astrophys.\ J.\ Lett.\ }{\rm #1}, {\rm#2} {\rm(#3)}}
\def\ajp#1,#2(#3){{\rm Am.\ J.\ Phys.\ }{\rm #1}, {\rm#2} {\rm(#3)}}
\def\apl#1,#2(#3){{\rm Appl.\ Phys. Lett.\ }{\rm #1}, {\rm#2} {\rm(#3)}}
\def\apny#1,#2(#3){{\rm Ann.\ Phys.\ (NY)\ }{\rm #1}, {\rm#2} {\rm(#3)}}
\def\apnyB#1,#2(#3){{\rm Ann.\ Phys.\ (NY)\ }{\rm B#1}, {\rm#2} {\rm(#3)}}
\def\apD#1,#2(#3){{\rm Ann.\ Phys.\ }{\rm D#1}, {\rm#2} {\rm(#3)}}
\def\ap#1,#2(#3){{\rm Ann.\ Phys.\ }{\rm #1}, {\rm#2} {\rm(#3)}}
\def\ass#1,#2(#3){{\rm Ap.\ Space Sci.\ }{\rm #1}, {\rm#2} {\rm(#3)}}
\def\astropp#1,#2(#3)%
    {{\rm Astropart.\ Phys.\ }{\rm #1}, {\rm#2} {\rm(#3)}}
\def\aap#1,#2(#3)%
    {{\rm Astron.\ \& Astrophys.\ }{\rm #1}, {\rm#2} {\rm(#3)}}
\def\araa#1,#2(#3)%
    {{\rm Ann.\ Rev.\ Astron.\ Astrophys.\ }{\rm #1}, {\rm#2} {\rm(#3)}}
\def\arnps#1,#2(#3)%
    {{\rm Ann.\ Rev.\ Nucl.\ and Part.\ Sci.\ }{\rm #1}, {\rm#2} {\rm(#3)}}
\def\arns#1,#2(#3)%
   {{\rm Ann.\ Rev.\ Nucl.\ Sci.\ }{\rm #1}, {\rm#2} {\rm(#3)}}
\def\cpc#1,#2(#3){{\rm Comp.\ Phys.\ Comm.\ }{\rm #1}, {\rm#2} {\rm(#3)}}
\def\cjp#1,#2(#3){{\rm Can.\ J.\ Phys.\ }{\rm #1}, {\rm#2} {\rm(#3)}}
\def\cmp#1,#2(#3){{\rm Commun.\ Math.\ Phys.\ }{\rm #1}, {\rm#2} {\rm(#3)}}
\def\cnpp#1,#2(#3)%
   {{\rm Comm.\ Nucl.\ Part.\ Phys.\ }{\rm #1}, {\rm#2} {\rm(#3)}}
\def\cnppA#1,#2(#3)%
   {{\rm Comm.\ Nucl.\ Part.\ Phys.\ }{\rm A#1}, {\rm#2} {\rm(#3)}}
\def\epjC#1,#2(#3){{\rm Eur.\ Phys.\ J.\ }{\rm C#1}, {\rm#2} {\rm(#3)}}
\def\el#1,#2(#3){{\rm Europhys.\ Lett.\ }{\rm #1}, {\rm#2} {\rm(#3)}}
\def\hpa#1,#2(#3){{\rm Helv.\ Phys.\ Acta }{\rm #1}, {\rm#2} {\rm(#3)}}
\def\ieeetNS#1,#2(#3)%
    {{\rm IEEE Trans.\ }{\rm NS#1}, {\rm#2} {\rm(#3)}}
\def\IEEE #1,#2(#3)%
    {{\rm IEEE }{\rm #1}, {\rm#2} {\rm(#3)}}
\def\ijar#1,#2(#3)%
  {{\rm Int.\ J.\ of Applied Rad.\ } {\rm #1}, {\rm#2} {\rm(#3)}}
\def\ijari#1,#2(#3)%
  {{\rm Int.\ J.\ of Applied Rad.\ and Isotopes\ } {\rm #1}, {\rm#2} {\rm(#3)}}
\def\jcap#1,#2(#3){{\rm JCAP\ }{\rm #1}, {\rm#2} {\rm(#3)}}
\def\jcp#1,#2(#3){{\rm J.\ Chem.\ Phys.\ }{\rm #1}, {\rm#2} {\rm(#3)}}
\def\jgr#1,#2(#3){{\rm J.\ Geophys.\ Res.\ }{\rm #1}, {\rm#2} {\rm(#3)}}
\def\jetp#1,#2(#3){{\rm Sov.\ Phys.\ JETP\ }{\rm #1}, {\rm#2} {\rm(#3)}}
\def\jetpl#1,#2(#3)%
   {{\rm Sov.\ Phys.\ JETP Lett.\ }{\rm #1}, {\rm#2} {\rm(#3)}}
\def\jhep#1,#2(#3){{\rm JHEP\ }{\rm #1}, {\rm#2} {\rm(#3)}}
\def\jpA#1,#2(#3){{\rm J.\ Phys.\ }{\rm A#1}, {\rm#2} {\rm(#3)}}
\def\jpG#1,#2(#3){{\rm J.\ Phys.\ }{\rm G#1}, {\rm#2} {\rm(#3)}}
\def\jpamg#1,#2(#3)%
    {{\rm J.\ Phys.\ A: Math.\ and Gen.\ }{\rm #1}, {\rm#2} {\rm(#3)}}
\def\jpcrd#1,#2(#3)%
    {{\rm J.\ Phys.\ Chem.\ Ref.\ Data\ } {\rm #1}, {\rm#2} {\rm(#3)}}
\def\jpsj#1,#2(#3){{\rm J.\ Phys.\ Soc.\ Jpn.\ }{\rm G#1}, {\rm#2} {\rm(#3)}}
\def\lnc#1,#2(#3){{\rm Lett.\ Nuovo Cimento\ } {\rm #1}, {\rm#2} {\rm(#3)}}
\def\nature#1,#2(#3){{\rm Nature} {\rm #1}, {\rm#2} {\rm(#3)}}
\def\nc#1,#2(#3){{\rm Nuovo Cimento} {\rm #1}, {\rm#2} {\rm(#3)}}
\def\nim#1,#2(#3)%
   {{\rm Nucl.\ Instrum.\ Methods\ }{\rm #1}, {\rm#2} {\rm(#3)}}
\def\nimA#1,#2(#3)%
    {{\rm Nucl.\ Instrum.\ Methods\ }{\rm A#1}, {\rm#2} {\rm(#3)}}
\def\nimB#1,#2(#3)%
    {{\rm Nucl.\ Instrum.\ Methods\ }{\rm B#1}, {\rm#2} {\rm(#3)}}
\def\np#1,#2(#3){{\rm Nucl.\ Phys.\ }{\rm #1}, {\rm#2} {\rm(#3)}}
\def\mnras#1,#2(#3){{\rm ASK.\ GEORGE.\ }{\rm #1}, {\rm#2} {\rm(#3)}}
\def\medp#1,#2(#3){{\rm Med.\ Phys.\ }{\rm #1}, {\rm#2} {\rm(#3)}}
\def\mplA#1,#2(#3){{\rm Mod.\ Phys.\ Lett.\ }{\rm A#1}, {\rm#2} {\rm(#3)}}
\def\npA#1,#2(#3){{\rm Nucl.\ Phys.\ }{\rm A#1}, {\rm#2} {\rm(#3)}}
\def\npB#1,#2(#3){{\rm Nucl.\ Phys.\ }{\rm B#1}, {\rm#2} {\rm(#3)}}
\def\npBps#1,#2(#3){{\rm Nucl.\ Phys.\ (Proc.\ Supp.) }{\rm B#1},
{\rm#2} {\rm(#3)}}
\def\pasp#1,#2(#3){{\rm Pub.\ Astron.\ Soc.\ Pac.\ }{\rm #1}, {\rm#2} {\rm(#3)}}
\def\pl#1,#2(#3){{\rm Phys.\ Lett.\ }{\rm #1}, {\rm#2} {\rm(#3)}}
\def\fp#1,#2(#3){{\rm Fortsch.\ Phys.\ }{\rm #1}, {\rm#2} {\rm(#3)}}
\def\ijmpA#1,#2(#3)%
   {{\rm Int.\ J.\ Mod.\ Phys.\ }{\rm A#1}, {\rm#2} {\rm(#3)}}
\def\ijmpE#1,#2(#3)%
   {{\rm Int.\ J.\ Mod.\ Phys.\ }{\rm E#1}, {\rm#2} {\rm(#3)}}
\def\plB#1,#2(#3){{\rm Phys.\ Lett.\ }{\rm B#1}, {\rm#2} {\rm(#3)}}
\def\pnasus#1,#2(#3)%
   {{\it Proc.\ Natl.\ Acad.\ Sci.\ \rm (US)}{B#1}, {\rm#2} {\rm(#3)}}
\def\ppsA#1,#2(#3){{\rm Proc.\ Phys.\ Soc.\ }{\rm A#1}, {\rm#2} {\rm(#3)}}
\def\ppsB#1,#2(#3){{\rm Proc.\ Phys.\ Soc.\ }{\rm B#1}, {\rm#2} {\rm(#3)}}
\def\pr#1,#2(#3){{\rm Phys.\ Rev.\ }{\rm #1}, {\rm#2} {\rm(#3)}}
\def\prA#1,#2(#3){{\rm Phys.\ Rev.\ }{\rm A#1}, {\rm#2} {\rm(#3)}}
\def\prB#1,#2(#3){{\rm Phys.\ Rev.\ }{\rm B#1}, {\rm#2} {\rm(#3)}}
\def\prC#1,#2(#3){{\rm Phys.\ Rev.\ }{\rm C#1}, {\rm#2} {\rm(#3)}}
\def\prD#1,#2(#3){{\rm Phys.\ Rev.\ }{\rm D#1}, {\rm#2} {\rm(#3)}}
\def\prept#1,#2(#3){{\rm Phys.\ Reports\ } {\rm #1}, {\rm#2} {\rm(#3)}}
\def\preptC#1,#2(#3){{\rm Phys.\ Reports\ } {\rm C#1}, {\rm#2} {\rm(#3)}}
\def\prslA#1,#2(#3)%
   {{\rm Proc.\ Royal Soc.\ London }{\rm A#1}, {\rm#2} {\rm(#3)}}
\def\prl#1,#2(#3){{\rm Phys.\ Rev.\ Lett.\ }{\rm #1}, {\rm#2} {\rm(#3)}}
\def\ps#1,#2(#3){{\rm Phys.\ Scripta\ }{\rm #1}, {\rm#2} {\rm(#3)}}
\def\ptp#1,#2(#3){{\rm Prog.\ Theor.\ Phys.\ }{\rm #1}, {\rm#2} {\rm(#3)}}
\def\ppnp#1,#2(#3)%
	{{\rm Prog.\ in Part.\ Nucl.\ Phys.\ }{\rm #1}, {\rm#2} {\rm(#3)}}
\def\ptps#1,#2(#3)%
   {{\rm Prog.\ Theor.\ Phys.\ Supp.\ }{\rm #1}, {\rm#2} {\rm(#3)}}
\def\pw#1,#2(#3){{\rm Part.\ World\ }{\rm #1}, {\rm#2} {\rm(#3)}}
\def\pzetf#1,#2(#3)%
   {{\rm Pisma Zh.\ Eksp.\ Teor.\ Fiz.\ }{\rm #1}, {\rm#2} {\rm(#3)}}
\def\rgss#1,#2(#3){{\rm Revs.\ Geophysics \& Space Sci.\ }{\rm #1},
        {\rm#2} {\rm(#3)}}
\def\rmp#1,#2(#3){{\rm Rev.\ Mod.\ Phys.\ }{\rm #1}, {\rm#2} {\rm(#3)}}
\def\rnc#1,#2(#3){{\rm Riv.\ Nuovo Cimento\ } {\rm #1}, {\rm#2} {\rm(#3)}}
\def\rpp#1,#2(#3)%
    {{\rm Rept.\ on Prog.\ in Phys.\ }{\rm #1}, {\rm#2} {\rm(#3)}}
\def\science#1,#2(#3){{\rm Science\ } {\rm #1}, {\rm#2} {\rm(#3)}}
\def\sjnp#1,#2(#3)%
   {{\rm Sov.\ J.\ Nucl.\ Phys.\ }{\rm #1}, {\rm#2} {\rm(#3)}}
\def\sjpn#1,#2(#3)%
   {{\rm Sov.\ J.\ Part.\ Nucl.\ }{\rm #1}, {\rm#2} {\rm(#3)}}
\def\panp#1,#2(#3)%
   {{\rm Phys.\ Atom.\ Nucl.\ }{\rm #1}, {\rm#2} {\rm(#3)}}
\def\spu#1,#2(#3){{\rm Sov.\ Phys.\ Usp.\ }{\rm #1}, {\rm#2} {\rm(#3)}}
\def\surveyHEP#1,#2(#3)%
    {{\rm Surv.\ High Energy Physics\ } {\rm #1}, {\rm#2} {\rm(#3)}}
\def\yf#1,#2(#3){{\rm Yad.\ Fiz.\ }{\rm #1}, {\rm#2} {\rm(#3)}}
\def\zetf#1,#2(#3)%
   {{\rm Zh.\ Eksp.\ Teor.\ Fiz.\ }{\rm #1}, {\rm#2} {\rm(#3)}}
\def\zp#1,#2(#3){{\rm Z.~Phys.\ }{\rm #1}, {\rm#2} {\rm(#3)}}
\def\zpA#1,#2(#3){{\rm Z.~Phys.\ }{\rm A#1}, {\rm#2} {\rm(#3)}}
\def\zpC#1,#2(#3){{\rm Z.~Phys.\ }{\rm C#1}, {\rm#2} {\rm(#3)}}

%% file: abstract.tex
\begin{abstract}
Quantum loops induce an anomaly, \amu, in the magnetic moment of the muon that can be 
accurately measured. Its Standard Model prediction is limited in precision by 
contributions from hadronic vacuum polarisation of the photon. The dominant 
lowest-order hadronic term can be calculated with a combination of experimental 
cross section data, involving \ee annihilation to hadrons, and perturbative QCD. 
These are used to evaluate an energy-squared dispersion integral that strongly 
emphasises low photon virtualities. The dominant contribution to the integral 
stems from the two-pion channel that can be measured both in \ee annihilation 
and in $\tau$ decays. The corresponding \ee and $\tau$-based predictions of 
\amu exhibit deviations by, respectively, $3.6\sigma$ and $2.4\sigma$ 
from experiment, leaving room for a possible interpretation in terms of new physics. 
This talk reviews the status of the Standard Model prediction with emphasis on 
the lowest-order hadronic contribution. Also given is the latest result for the 
running electromagnetic fine structure constant at the $Z$-mass pole, whose precision
is limited by hadronic vacuum polarisation contributions, determined in a 
way similar to those of the magnetic anomaly. 
\end{abstract}

%% file: bulk.tex
\section{INTRODUCTION}

The Dirac equation predicts a muon magnetic moment, 
${\bf M}=g_{\mu}\frac{e}{2m_{\mu}}{\bf S}$, with gyromagnetic ratio
$g_{\mu}=2$. Quantum loop effects lead to a small calculable deviation
from $g_{\mu}=2$, parametrised by the anomalous magnetic moment
\beq
   \amu \equiv \frac{g_{\mu}-2}{2}\,.
\eeq
\noindent
That quantity can be accurately measured and, within the Standard Model (SM) 
framework, precisely predicted. Hence, comparison of experiment and theory 
tests the SM at its quantum loop level. A deviation in $a^{\rm exp}_{\mu}$ 
from the SM expectation would signal
effects of new physics, with current sensitivity reaching up to mass
scales of ${\cal O} ({\rm TeV})$~\cite{czarnecki2,davier}.\footnote
{ 
   Although the corresponding electron anomalous magnetic moment has been 
   measured approximately 800 times more accurately than the muon one, its 
   sensitivity to new physics is expected to be about 50 times lower, owing 
   to the quadratic lepton-mass dependence of the virtual contribution from 
   new, heavy particles. 
}
For recent and very thorough muon $g-2$ reviews, see Refs.~\cite{miller,jn}.

\begin{figure*}[t]
\newcommand\diagramScale{3cm}
\begin{center}
  \includegraphics[width=\diagramScale]{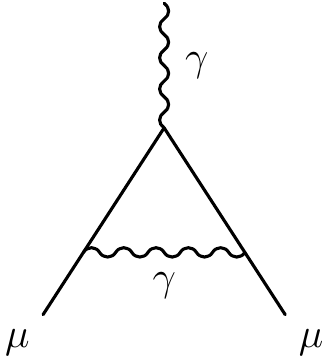}\hspace{0.7cm}
  \includegraphics[width=\diagramScale]{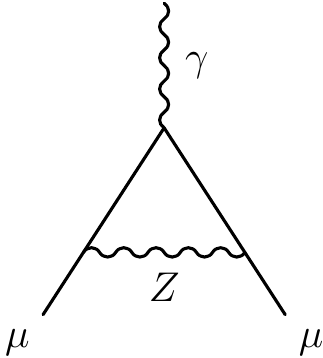}\hspace{0.7cm}
  \includegraphics[width=\diagramScale]{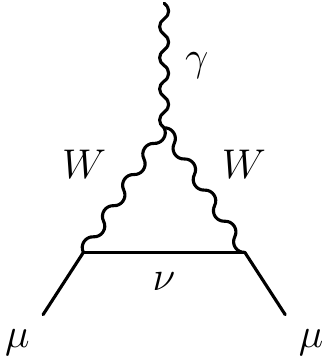}\hspace{0.7cm}
  \includegraphics[width=\diagramScale]{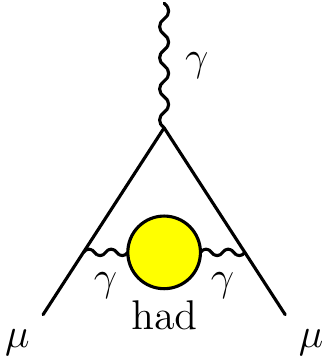}
\end{center}
\vspace{-0.4cm}
\caption{Representative diagrams contributing to $a^{\rm SM}_\mu$.
         From left to right: first order QED (Schwinger term),
         lowest-order weak, lowest-order hadronic.}
\label{fig:g-2feyn}
\end{figure*}
The E821 experiment at Brookhaven National Lab
(BNL) studied the spin precession of $\mu^{+}$ and $\mu^{-}$ in a constant
external magnetic field as they circulated in a confining storage ring.
The muon momentum of $3.1\,\gev$ was chosen such that the difference between 
precession and cyclotron frequency, $\omega_a$, is approximately independent 
of the electrical quadrupole field, required for the vertical focusing of the 
muons. This is the case for a Lorentz factor of $\gamma_\mu=\sqrt{1+1/\amu}\approx29.3$.
The muon anomalous magnetic moment is then directly given by $\amu=\omega_a m_\mu c/(eB)$.
The magnetic field strength and homogeneity was precisely measured from NMR probes
frequently pulled with a trolley on rails through the ring, and $\omega_a$ was obtained 
from a fit 
to the decay electron (or positron) counting rates in scintillators installed along 
the cyclotron. The results were obtained by independently blinding $B$ and $\omega_a$  
until all systematic studies were terminated to satisfaction. The final published results 
read~\cite{bennett}\footnote
{
   The original results reported by the experiment 
   have been updated in Eqs.~(\ref{eq:amu_exp_nump})--(\ref{eq:amu_exp_num}) to 
   the newest value for the absolute muon-to-proton magnetic ratio  
   $\lambda=3.183345137 \pm 85$~\cite{codata}. The change induced in 
   $a^{\rm exp}_{\mu}$ with respect to the value of 
   $\lambda=3.18334539 \pm 10$ used in Ref.\cite{bennett} amounts 
   to $+0.92\times10^{-10}$~\cite{pdgg-2rev}.
}
\beqn
\label{eq:amu_exp_nump}
   a^{\rm exp}_{\mu+} & =&  (11\,659\,204 \pm 6 \pm 5) \times 10^{-10}\,, \\
\label{eq:amu_exp_numm}
   a^{\rm exp}_{\mu-} & =&  (11\,659\,215 \pm 8 \pm 3) \times 10^{-10}\,,
\eeqn
where the first errors are statistical and the second systematic.
Assuming \CPT invariance and taking into account correlations
between systematic errors, one finds for their average~\cite{bennett}
\beq
\label{eq:amu_exp_num}
   a^{\rm exp}_{\mu}=(11\,659\,208.9 \pm 5.4 \pm 3.3) \times 10^{-10}~.
\eeq
These results represent about a factor of 14 improvement over the
classic CERN experiments of the 1970's~\cite{cern70}.

The SM prediction for $a^{\rm SM}_{\mu}$ is conveniently separated into
three parts (see Fig.~\ref{fig:g-2feyn} for representative Feynman
diagrams)
\beq
    a^{\rm SM}_{\mu}=a^{\rm QED}_{\mu}+ a^{\rm EW}_{\mu} + \amuhad~.
\eeq
The QED part includes all photonic and leptonic $(e, \mu,\tau)$ loops
starting with the classic $\alpha/2\pi$ Schwinger contribution. It has
been computed through 4 loops and estimated at the 5-loop
level~\cite{kinoshita}
\beqn
\label{eq:amu_qed}
\lefteqn{
    a^{\rm QED}_{\mu} = \frac{\alpha}{2\pi} +
                0.765857410(27)\left(\frac{\alpha}{\pi}\right)^{\!2}} \nonumber\\
                && +\;24.05050964(43)\left(\frac{\alpha}{\pi}\right)^{\!3} \\
                && +\; 130.8055(80)\left(\frac{\alpha}{\pi}\right)^{\!4} +
                663(20)\left(\frac{\alpha}{\pi}\right)^{\!5} +
                \cdots\,,\nonumber
\eeqn
where the errors in each term are given in parentheses.
Employing $\alpha^{-1} = 137.035999084(51)$, determined~\cite{kinoshita,newalpha}
from the electron $a_{e}$ measurement, leads~to
\beq
\label{eq:amu_qed_num}
   a^{\rm QED}_{\mu} = (116\,584\,718.09 \pm 0.15) \times 10^{-11}\,,
\eeq
where the error account for the uncertainties in the coefficients~(\ref{eq:amu_qed})
and in $\alpha$. 

Loop contributions involving heavy $W^{\pm},Z$ or Higgs particles are collectively 
labelled as $a^{\rm EW}_{\mu}$. They are suppressed by at least a factor of
$\frac{\alpha}{\pi}\frac{m^{2}_{\mu}}{m^{2}_{W}}\simeq4\times 10^{-9}$.
At 1-loop order one finds~\cite{jackiw}
\beqn
\label{eq:amu_ew_1loop}
   a^{\rm EW}_{\mu}[\hbox{1-loop}] & =&
       \frac{G_{\mu}m^{2}_{\mu}}{8\sqrt{2} \pi^{2}} 
        \Bigg[ \frac{5}{3}
       +\frac{1}{3} \left(1-4\sin^{2}\!\theta _{\rm W} \right)^{2}\nonumber\\
        &&+\; {\cal O} 
        \left( \frac{m^{2}_{\mu}} {M^{2}_{W}} \right)
        + {\cal O} \left( \frac {m^{2}_{\mu}}{m^{2}_{H}} \right)
        \Bigg]\,, \nonumber\\
 	    &=&  194.8 \times 10^{-11}\,,
\eeqn
for $\sin^{2}\!\theta_{\rm W}\equiv 1- M^{2}_{W}/M^{2}_{Z}\simeq 0.223$, and 
where $G_{\mu}\simeq1.166\times10^{-5}\;{\rm GeV}^{-2}$ is the Fermi coupling 
constant. Two-loop corrections are relatively large and negative~\cite{czarnecki}
\beq
\label{eq:amu_ew_2loop}
 a^{\rm EW}_{\mu}[\hbox{2-loop}] = (-40.7\pm1.0\pm1.8) \times 10^{-11}\,,
\eeq
where the errors stem from quark triangle loops and the assumed Higgs
mass range between 100 and $500\;{\rm GeV}$. The 3-loop leading
logarithms are negligible~\cite{czarnecki,degrassi}, ${\cal O}(10^{-12})$,
implying in total
\beq
\label{eq:amu_ew_num}
    a^{\rm EW}_{\mu} = (154\pm1\pm2) \times 10^{-11}~.
\eeq
\begin{figure*}[t]
\includegraphics[width=\figsize]{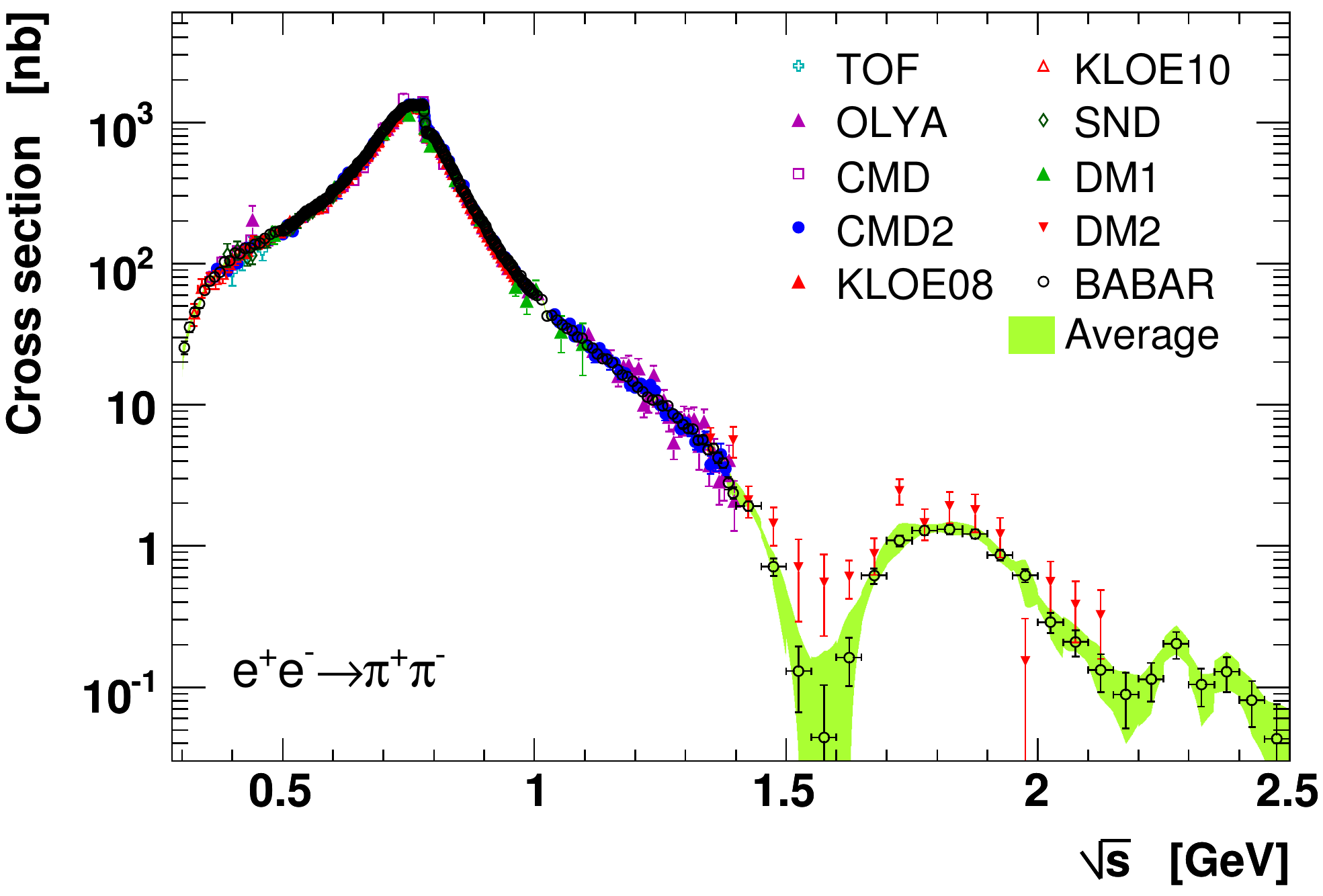}\hspace{\fighspace}
\includegraphics[width=\columnwidth]{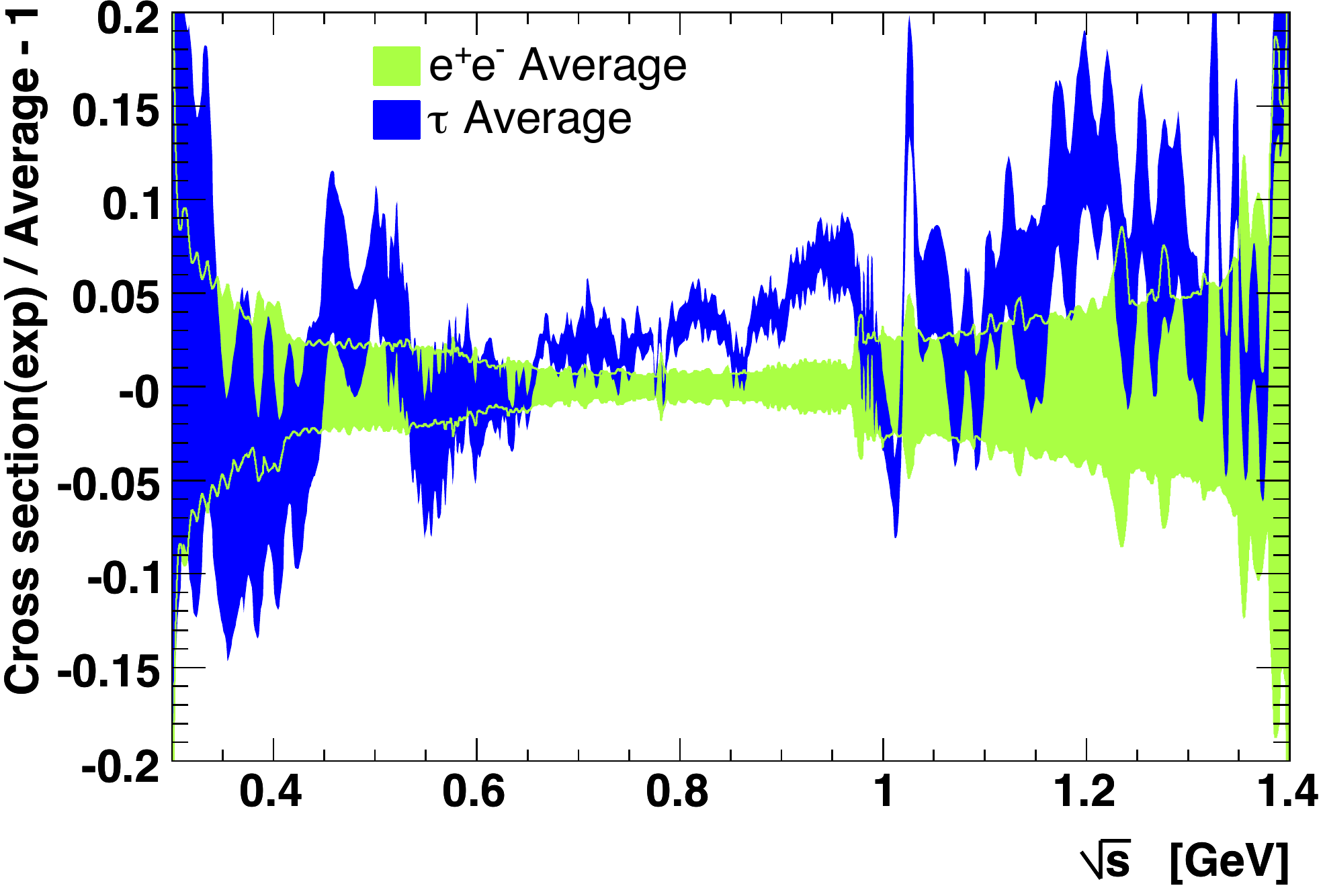}
\vspace{-0.1cm}
\caption[.]{Left panel: cross section of $\ee\to\pp$ versus centre-of-mass energy for different 
            energy ranges (see~\cite{dhmz2010} for references). The error bars show statistical 
            and systematic errors added in quadrature. The light shaded (green) band indicates
            the average within $1\,\sigma$ errors. 
            Right panel: relative difference between the average $\taum\to\pipiz\nut$ and 
            $\ee\to\pp$ cross sections. The individual measurements show agreement 
            between Belle/CLEO ($\tau$) and BABAR ($\ee$), but discrepancies
            between $\tau$ and KLOE data.}
\label{fig:pipixsec}
\end{figure*}
\begin{figure*}[t]
\begin{center}
\includegraphics[width=\figsize]{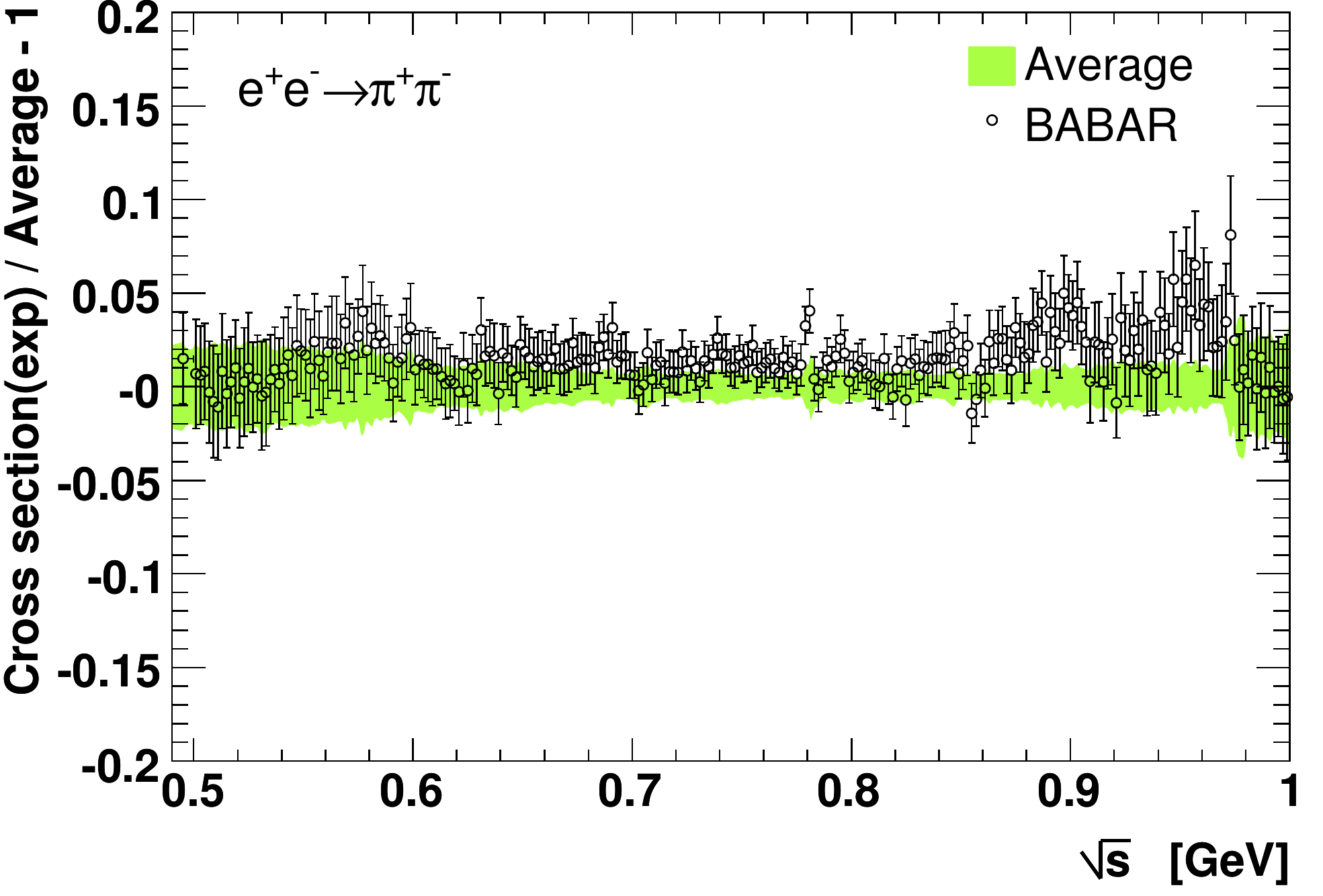}\hspace{\fighspace}
\includegraphics[width=\figsize]{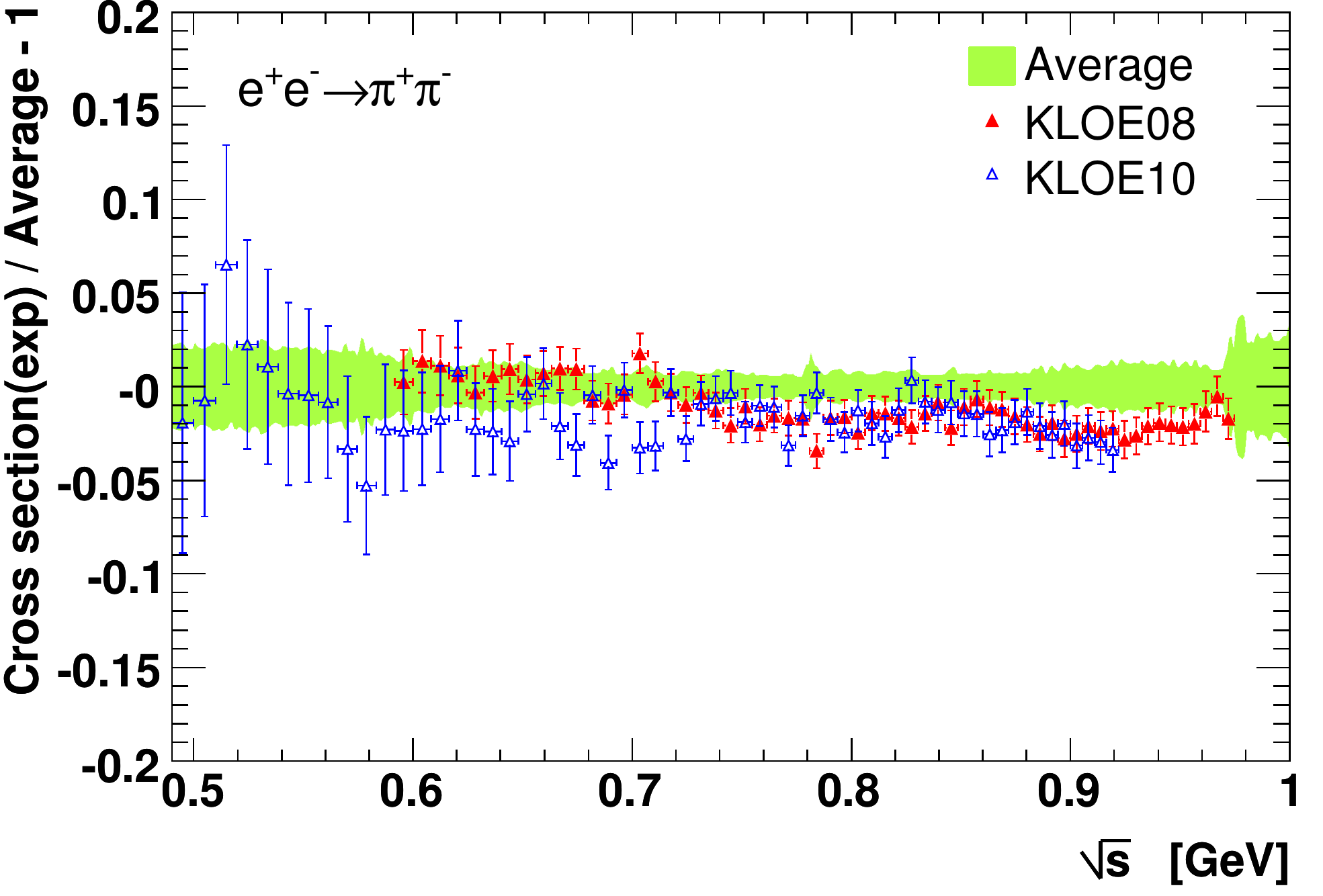}
\vspace{0.2cm}

\includegraphics[width=\figsize]{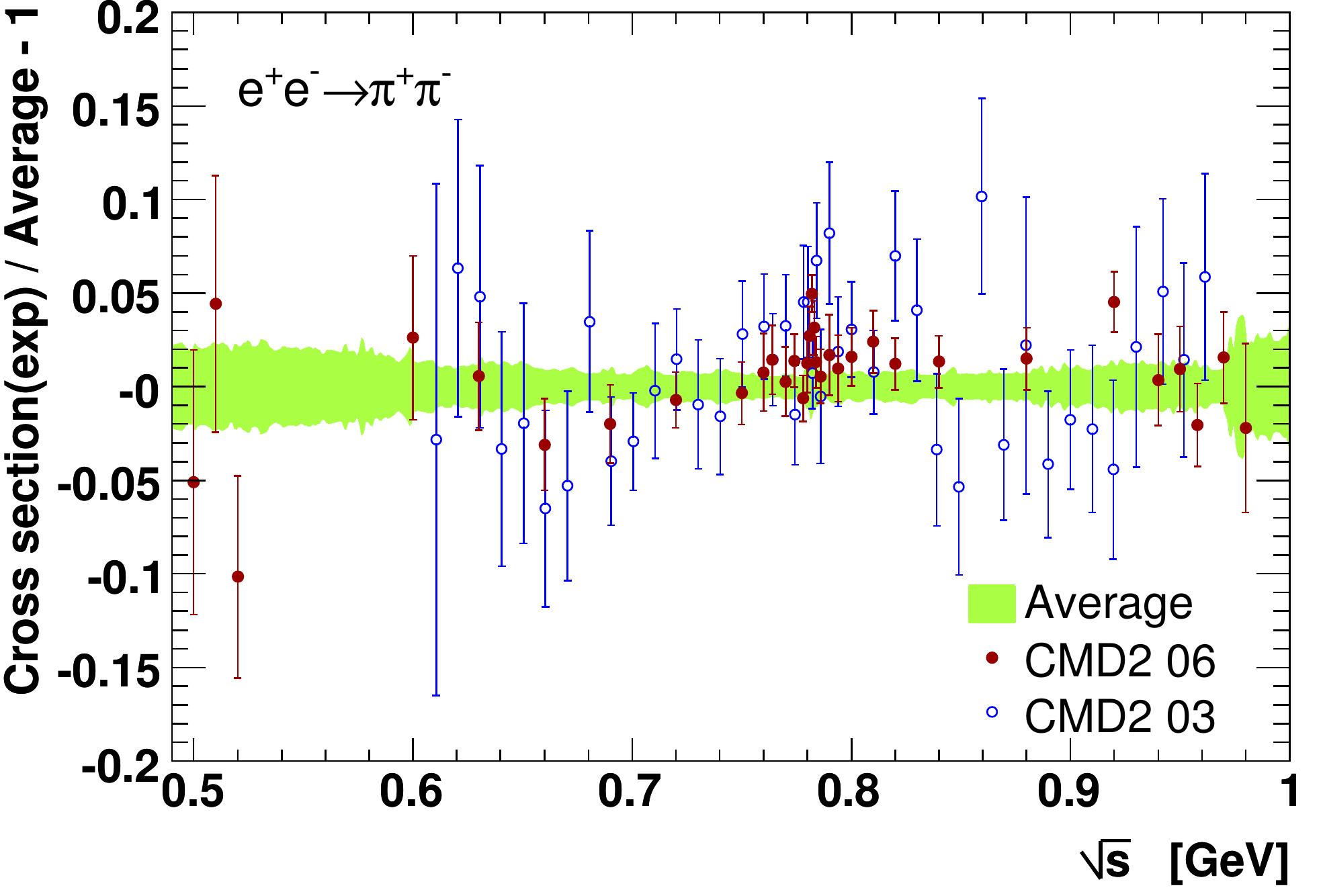}\hspace{\fighspace}
\includegraphics[width=\figsize]{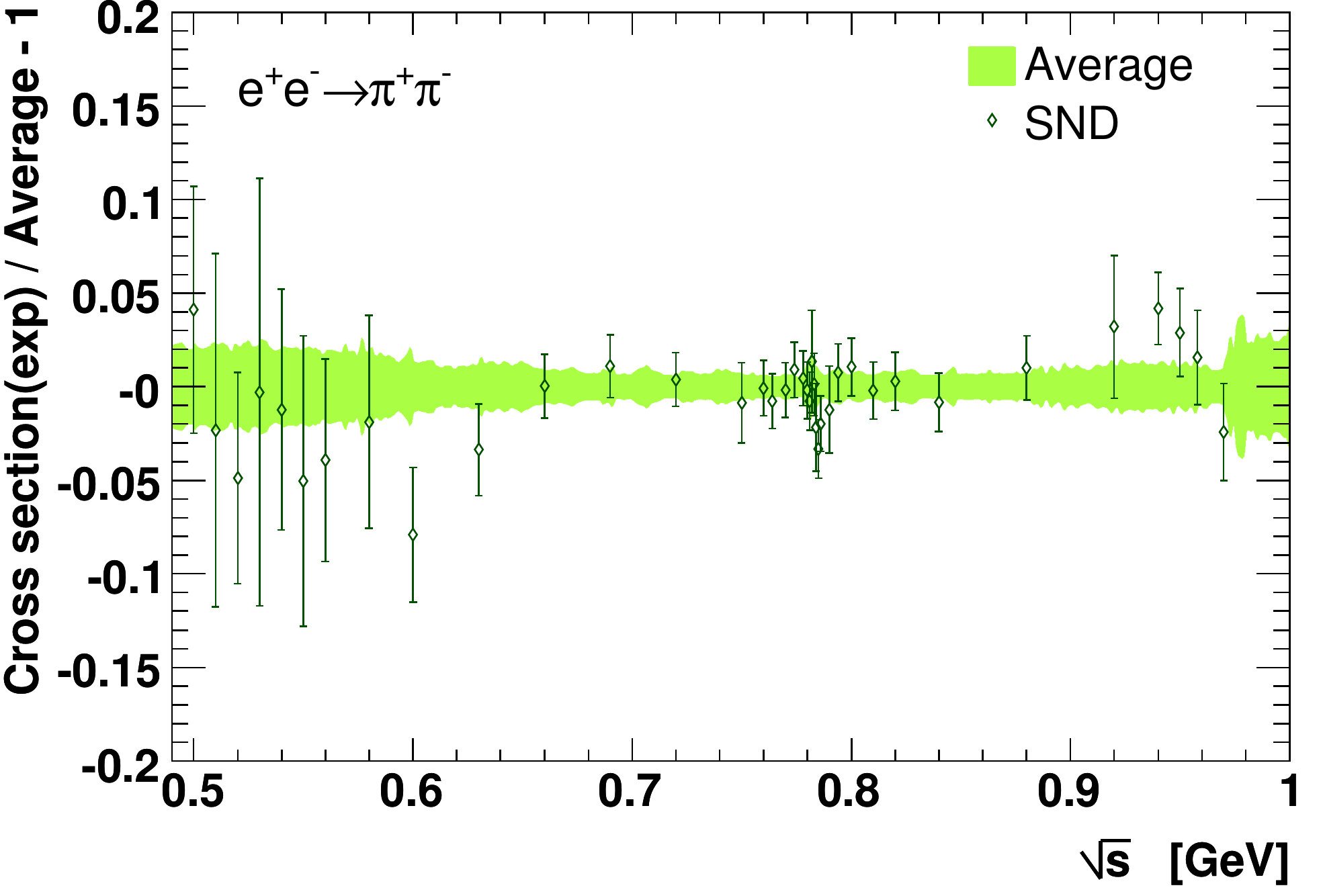}
\end{center}
\vspace{-0.4cm}
\caption[.]{ 
            Comparison between individual $\ee\to\pp$ cross section measurements from 
            BABAR~\cite{babarpipi}, KLOE\,08~\cite{kloe08}, KLOE\,10~\cite{kloe10},
            CMD2\,03~\cite{cmd203}, CMD2\,06~\cite{cmd2new}, SND~\cite{snd}, and 
            the average. The error bars show statistical and systematic 
            errors added in quadrature. 
}
\label{fig:comppipi}
\end{figure*}
Hadronic (quark and gluon) loop contributions to $a^{\rm SM}_{\mu}$
give rise to its main uncertainty. At present, those
effects are not calculable from first principles, but such an approach,
at least partially, may become possible as lattice QCD matures. Instead, one 
currently relies on a dispersion relation approach to evaluate the dominant 
lowest-order ${\cal O}(\alpha^2)$ hadronic vacuum polarisation 
contribution $\amuhadLO$ from corresponding cross section measurements or, 
where applicable, from perturbative QCD~\cite{rafael}
\beq
\label{eq:amu_had_lo}
 \amuhadLO =
           \frac{1}{3}\left(\frac{\alpha}{\pi}\right)^{\!\!2}\!
           \int\limits_{m_{\piz\gamma}^2}^\infty\!\!ds\,\frac{K(s)}{s}R^{(0)}(s)\,,
\eeq
where $K(s)$ is a QED kernel function~\cite{rafael2}, and where
$R^{(0)}(s)$ denotes the ratio of the bare\footnote
{
   The bare cross section is defined as the measured cross section
   corrected for initial-state radiation, electron-vertex loop
   contributions and vacuum-polarisation effects 
	in the photon propagator. QED effects in the
   hadron vertex and final state, as photon radiation, are included,
   \ie, not corrected. 
}
cross section for \ee annihilation into hadrons to the pointlike
muon-pair cross section at centre-of-mass energy $\sqrt{s}$. The function
$K(s)\sim1/s$ in Eq.~(\ref{eq:amu_had_lo}) gives a strong weight to the
low-energy part of the integral so that $\amuhadLO$ is dominated by the 
contribution from the $\rho(770)\to\pi\pi$ resonance. Equation~(\ref{eq:amu_had_lo})
is solved by using sums of exclusive cross section data at low centre-of-mass
energies (often chosen to be below $1.8\,\gev$), inclusive hadronic cross 
section data in the $c\cbar$ threshold region, and perturbative QCD elsewhere. 

A huge effort over 20 years and more by experimentalists and theorists 
went into the determination of the lowest-order hadronic contribution. 
The most significant improvements came from the experimental side with the 
availability of more accurate \ee cross section data from Novosibirsk, 
and by exploiting the high statistics data samples of the $B$ and $\Phi$ 
factories using the technique of radiative return. Using isospin symmetry,
precise hadronic $\tau$ decay data could also be used to complement the \ee 
data. The understanding that perturbative QCD works seamlessly down
to unexpectedly low energy scales, led to more extensive use of theory
to replace less precise data. 

The analysis of the hadronic contribution reported here has been 
published~\cite{dhmz2010} after the conference.
It includes new \pp cross-section data from KLOE, all 
available multi-hadron data from BABAR, a reestimation 
of missing low-energy contributions using results on cross sections and 
process dynamics from BABAR, a reevaluation of all experimental contributions 
using newly developed software, and a reanalysis of inter-experiment
and inter-channel correlations, and finally a reevaluation of the continuum 
contributions from perturbative QCD at four loops. 

\section{New hadronic cross section data}

\begin{figure}[t]
\includegraphics[width=\figsize]{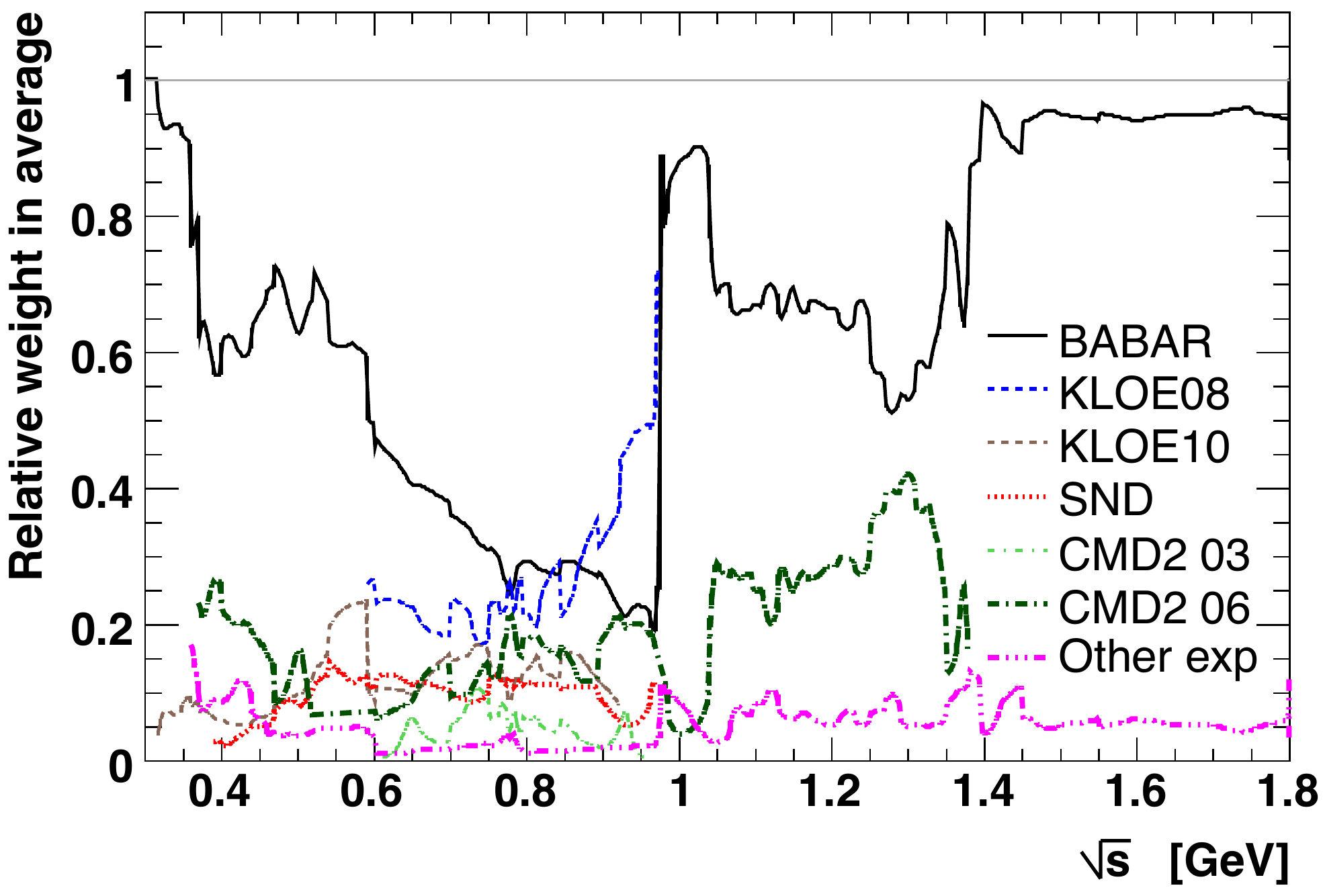}
\vspace{-0.4cm}
\caption[.]{Relative local averaging weight per experiment versus centre-of-mass energy 
            in $\ee\to\pp$. }
\label{fig:weights}
\end{figure}
\begin{figure*}[t]
\begin{center}
\includegraphics[width=\figsize]{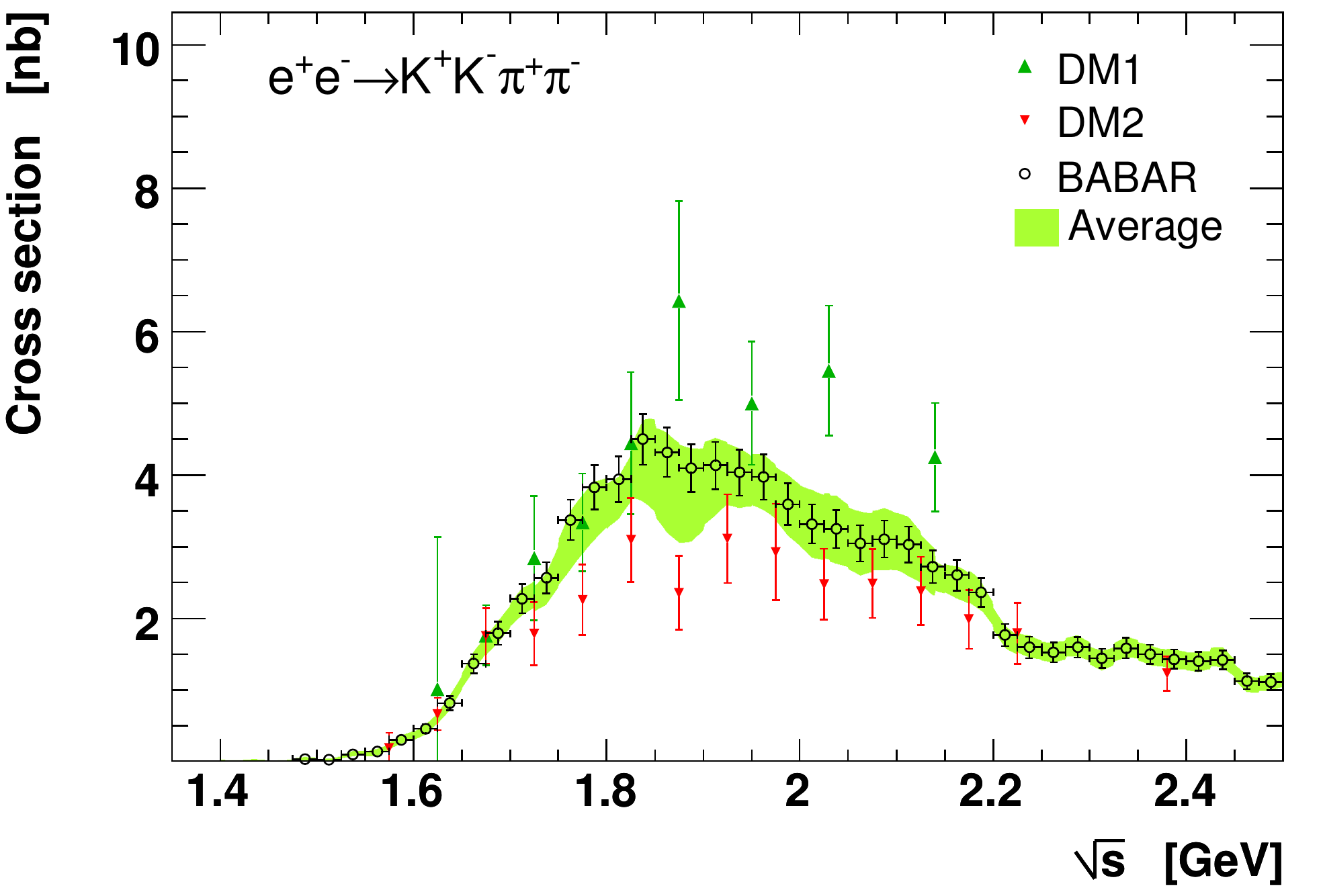}\hspace{\fighspace}
\includegraphics[width=\figsize]{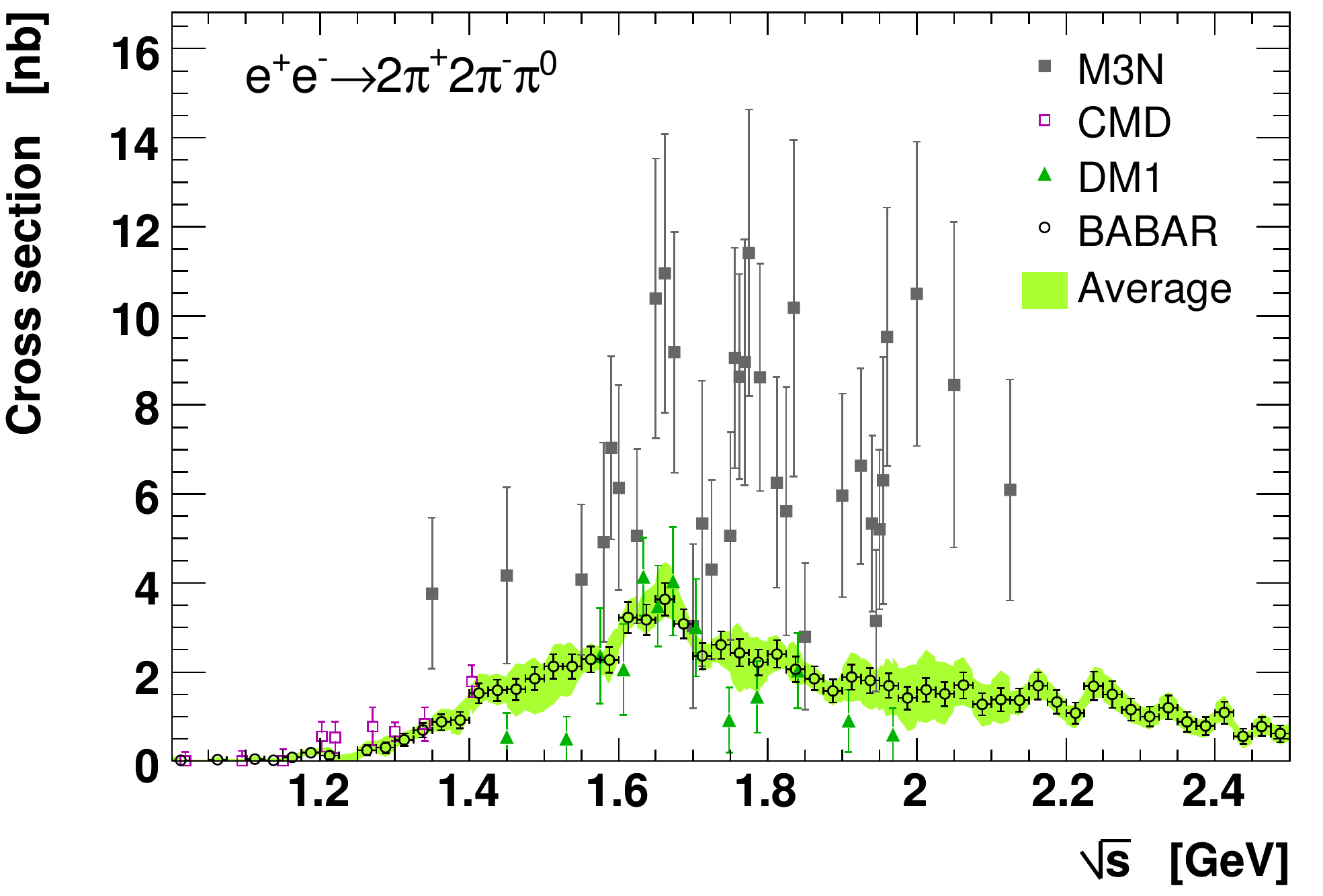}
\vspace{0.2cm}

\includegraphics[width=\figsize]{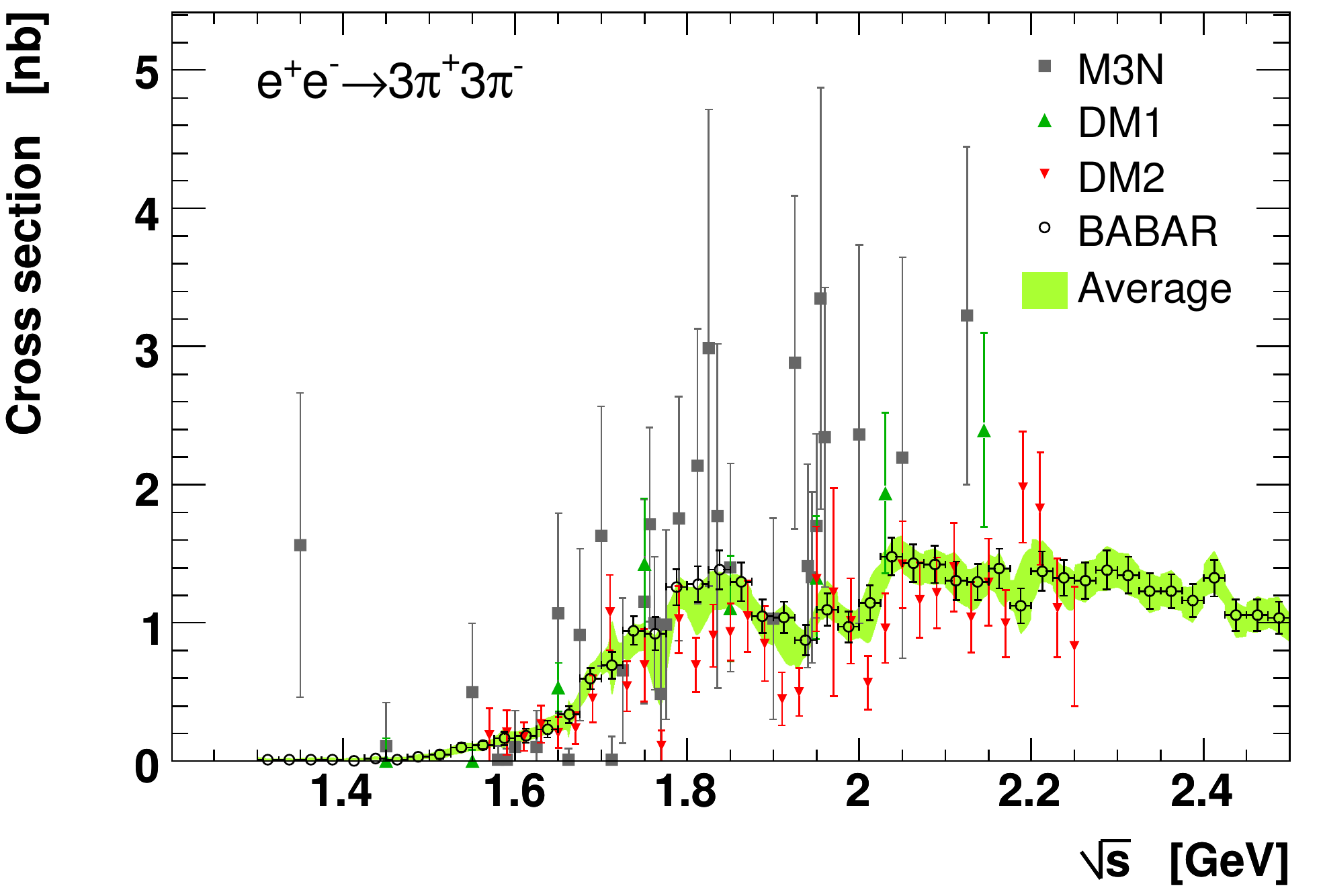}\hspace{\fighspace}
\includegraphics[width=\figsize]{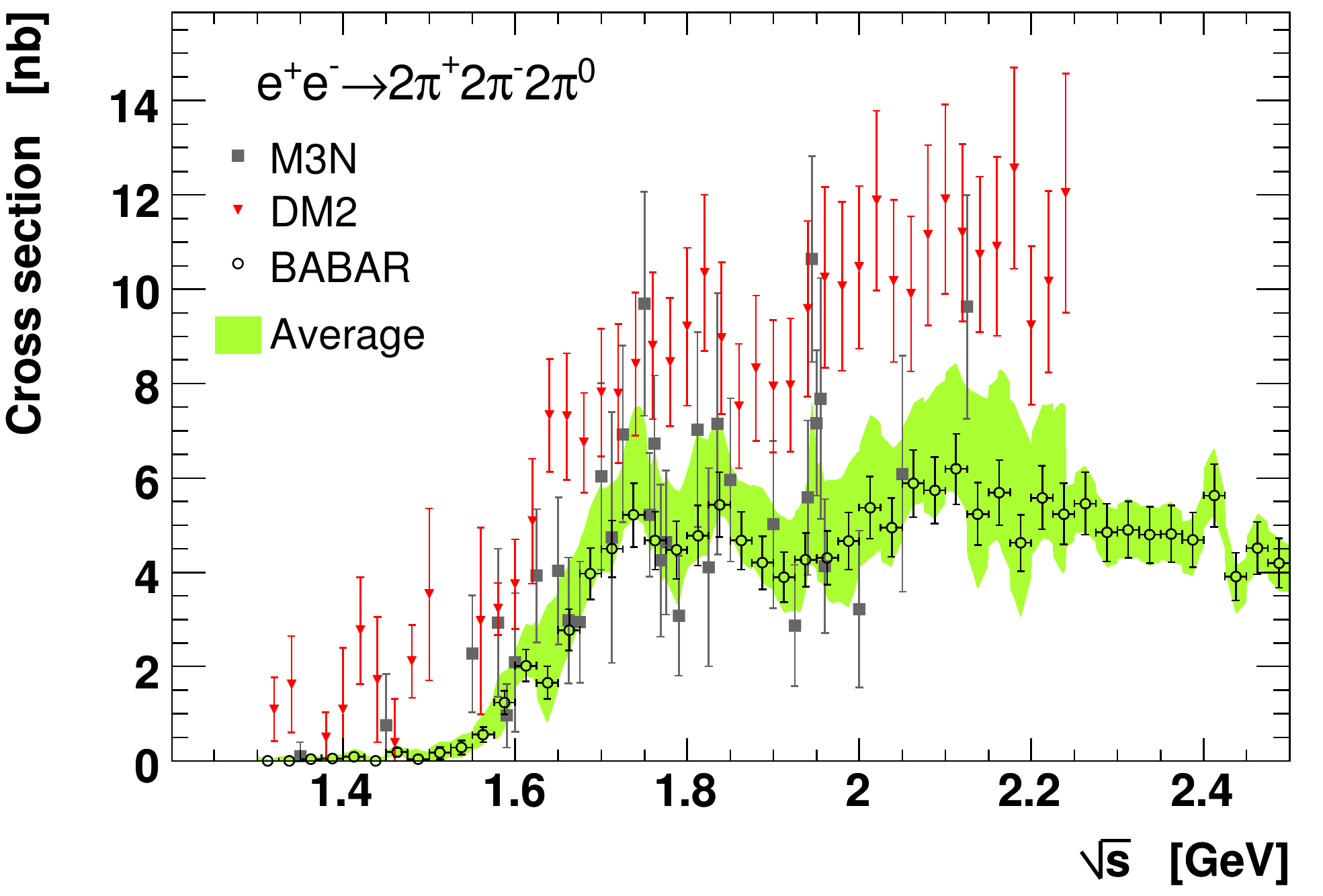}
\end{center}
\vspace{-0.5cm}
\caption[.]{ 
            Cross section data for the final states $\Kp\Km\pp$ (upper left),
            $2\pip2\pim\piz$ (upper right), $3\pip3\pim$ (lower left) and
            $2\pip2\pim2\piz$ (lower right). 
            The shaded (green) bands give the averages within $1\,\sigma$ errors, 
            locally rescaled in case of incompatibilities. 
            The BABAR data points are taken from Refs.~\cite{babar4pi,babar4pipi0,babar6pi}.
            The other data points are referenced in~\cite{dhmz2010}.
}
\label{fig:xsecmulti}
\end{figure*}
The KLOE Collaboration has published new $\pp\gamma$ cross section data 
with \pp invariant  mass-squared between 0.1 and $0.85\,\gev^2$~\cite{kloe10}. 
The radiative photon in this analysis is required to be detected in the 
electromagnetic calorimeter, which reduces the selected data sample to 
events with large photon scattering angle. The new data are found to 
be in agreement with, but less precise than previously 
published data using small angle photon scattering~\cite{kloe08}. They
exhibit a discrepancy, on the $\rho$ resonance peak and above, with other 
\pp data, in particular those from BABAR, obtained with the use of the same ISR 
technique~\cite{babarpipi}, and with data from $\taum\to\pipiz\nut$ 
decays~\cite{eetaunew} (\cf right-hand panel of Fig.~\ref{fig:pipixsec}). 

Figure~\ref{fig:pipixsec} (left) shows the available $\ee\to\pp$ cross section 
measurements in various panels for different centre-of-mass energies ($\sqrt{s}$). The 
light shaded (green) band indicates the average within $1\,\sigma$ errors.
The deviation between the average and the most precise individual measurements is 
depicted in Fig.~\ref{fig:comppipi}. Figure~\ref{fig:weights}
shows the weights versus $\sqrt{s}$ the different experiments obtain in the locally 
performed average. BABAR and KLOE dominate the average over the entire energy range. 
Owing to the sharp radiator function, the available statistics for KLOE increases 
towards the $\phi$ mass, hence outperforming BABAR above $\sim$$0.8\,\gev$. For example, 
at $0.9\,\gev$ KLOE's small photon scattering angle data~\cite{kloe08} have statistical 
errors of $0.5\%$, which is twice smaller than that of BABAR (renormalising
BABAR to the 2.75 times larger KLOE bins at that energy). Conversely, at $0.6\,\gev$ the 
comparison reads $1.2\%$ (KLOE) versus $0.5\%$ (BABAR, again given in KLOE bins which 
are about 4.2 times larger than for BABAR at that energy). The discrepancy between the 
BABAR and KLOE data sets above $0.8\,\gev$ causes error rescaling in their 
average, and hence loss of precision. The group of experiments labelled ``other exp'' 
in Fig.~\ref{fig:weights} corresponds to older data with incomplete radiative 
corrections. Their weights are small throughout the entire energy domain. 

For the analysis in Ref.~\cite{dhmz2010} the contributions from the $\omega(782)$ 
and $\phi(1020)$ resonances were computed for the first time directly from the 
corresponding partial measurements in the $\pp\piz$, $\piz\gamma$, $\eta\gamma$, 
$\Kp\Km$, $\KS\KL$ channels. Small remaining decay modes were considered separately. 

Also included in Ref.~\cite{dhmz2010} are new, preliminary, $\ee\to\pp 2\piz$ cross 
section measurements from BABAR~\cite{babar2pi2pi0}, which significantly help
to constrain a contribution with disparate experimental information.\footnote
{\label{ftn:bcvc4pion}The new measurements also improve the conserved vector current 
   (CVC) predictions for the corresponding $\tau$ decays with four pions in the 
   final state. Reference~\cite{dhmz2010} finds
   $\BR_{\rm CVC}(\taum\to\pim3\piz\nut)=(1.07\pm0.06)\%$, to be compared to 
   the world average of the direct measurements $(1.04\pm0.07)\%$~\cite{pdg10}, 
   and $\BR_{\rm CVC}(\taum\to2\pim\pip\piz\nut)=(3.79\pm0.21)\%$, to be compared to 
   the direct measurement $(4.48\pm0.06)\%$. The deviation between prediction 
   and measurement in the latter channel amounts to $3.2\,\sigma$, compared to 
   $3.6\,\sigma$ without the BABAR data~\cite{dehz02}. 
}  

Precise BABAR data~\cite{babar4pipi0,babar6pi,babarkkpi,babarkkpipi} are
available for several higher multiplicity modes with and without kaons, which 
greatly benefit from the excellent particle identification capabilities 
of the BABAR detector, The new data help to discriminate between older, less precise
and sometimes contradicting measurements. Figure~\ref{fig:xsecmulti} shows the 
cross section measurements and averages for the channels $\Kp\Km\pp$ 
(upper left), $2\pip2\pim\piz$ (upper right), $3\pip3\pim$ (lower left), 
and $2\pip2\pim2\piz$ (lower right). The BABAR data supersede much less precise 
measurements from M3N, DM1 and DM2. In several occurrences, these older measurements 
overestimate the cross sections in comparison with BABAR, which contributes to 
the reduction in the size of the present evaluation of hadronic loop effects. 

Good agreement is observed among the measurements of the charm resonance 
region above the opening of the $D\Dbar$ channel~\cite{dhmz2010}.

Several five and six-pion modes involving \piz's, as well as $K\Kbar [n\pi]$ 
final states are still unmeasured. Their contributions are estimated from those 
of known channels by means of Pais classification of $N$-pion states with total 
isospin $I=0,1$~\cite{pais}. The new BABAR cross section data and 
results on process dynamics thereby allow more stringent constraints of the 
unknown contributions than the ones obtained in previous analyses~\cite{dehz02,dehz03}.
The reanalysis of the available information led to the following approaches~\cite{dhmz2010}:
\bei

\item {\em 5-pion channels:} isospin constrains the unmeasured $3\piz$ mode by  
      $\sigma (2\pi^+2\pi^-\piz) = 2\cdot\sigma (\pi^+\pi^-3\piz)$. The isospin-breaking
      (IB) $\eta\pi\pi$ contribution is subtracted and treated apart.

\item {\em 6-pion channels:} the unknown $\sigma(\pi^+\pi^-4\piz)$ is determined from 
      $\sigma(3\pi^+3\pi^-)$ and $\sigma (2\pi^+2\pi^-2\piz)$, together with an upper 
      limit from $\tau\to 6\pi\nu$ data on two unconstrained Pais partitions, and using
      the experimental fact that these modes are dominated by $\omega3\pi$. 
      IB contributions from $\eta\omega$ are subtracted and treated separately.

\item {\em KK$(n\pi)$ channels:} the missing modes $\KS\KL\piz$ and $\KS K^+\pi^-\piz$, 
      $\KS\KS\pi^+\pi^-$, $\KS\KL\pi^+\pi^-$ are estimated from $I=0,1$ isospin relations 
      using $K^\star(890) K$ dominance, and correcting for small $\phi\pi$ and $KK\rho$ 
      contributions. The $KK\pi\pi\pi$ contribution is determined from the measured
      $K^+K^-\pi^+\pi^-\piz$ using the observed $K^+K^-\omega$ dominance.

\item {\em $\eta4\pi$ channels:} the total contribution is estimated as twice the measured 
      $\sigma (\eta2\pi^+2\pi^-)$.

\end{itemize}
Conservative systematic errors are applied where the dynamical information is incomplete
to fully determine all contributing Pais partitions. 

\section{Data averaging and integration}

For the evaluation in Ref.~\cite{dhmz2010} all experimental cross section data 
used in the compilation have been evaluated with the software package HVPTools. It 
replaces linear interpolation between adjacent data points (``trapezoidal rule'') by 
quadratic interpolation, which is found from pseudo-model analyses, with known truth 
integrals, to be more accurate. The interpolation functions are locally averaged 
between experiments, whereby correlations between measurement points of the same 
experiment and among different experiments due to common systematic errors are 
fully taken into account. Incompatible measurements lead to error rescaling in 
the local averages, using the PDG prescription~\cite{pdg10}. 

The errors in the average and in the integration for each channel are obtained 
from large samples of pseudo Monte Carlo experiments, by fluctuating all data points 
within errors respecting their correlations. The integrals of the exclusive channels 
are then summed up, and the error of the sum is obtained by adding quadratically 
(linearly) all uncorrelated (correlated) errors.

Common sources of systematic errors also occur between measurements of different 
final state channels and must be taken into account when summing up the exclusive 
contributions. Such correlations mostly arise from luminosity uncertainties, if 
the data stem from the same experimental facility, and from radiative corrections. 
In total eight categories of correlated systematic uncertainties are distinguished~\cite{dhmz2010}.
Among those the most significant belong to radiative corrections, which are the same 
for CMD2 and SND, as well as to luminosity determinations by BABAR, CMD2 and SND 
(correlated per experiment for different channels, but independent between different 
experiments). 

\section{Results}

\begin{table}[t]
\caption[.]{Contributions to \amuhadLO (middle column) from the individual \pp cross section 
   measurements by BABAR~\cite{babarpipi}, KLOE~\cite{kloe10,kloe08}, 
   CMD2~\cite{cmd203,cmd2new}, and SND~\cite{snd}. Also given are the corresponding CVC predictions
   of the $\tau\to\pim\piz\nut$ branching fraction (right column), corrected for isospin-breaking
   effects~\cite{eetaunew}. Here the first error is experimental and the second estimates the 
   uncertainty in the isospin-breaking corrections. The predictions are to be compared 
   with the world average of the 
   direct branching fraction measurements $(25.51 \pm 0.09)\%$~\cite{pdg10}.
   For each experiment, all available data in the energy range from threshold 
   to $1.8\,\gev$ ($m_\tau$ for $\BR_{\rm CVC}$) are used, and the missing part is 
   completed by the combined \ee data. The corresponding (integrand dependent) 
   fractions of the full integrals provided by a given experiment are given in 
   parentheses.
   \label{tab:resultspipi}
}
\setlength{\tabcolsep}{0.0pc}
{\small
\begin{tabularx}{\columnwidth}{@{\extracolsep{\fill}}lrr} 
\hline\noalign{\smallskip}
Exp. & \mc{1}{c}{$\amuhadLO~[10^{-10}]$} & \mc{1}{c}{$\BR_{\rm CVC}~[\%]$} \\
\noalign{\smallskip}\hline\noalign{\smallskip}
BABAR & $514.1\pm 3.8~ (1.00)$  & $ 25.15 \pm 0.18 \pm 0.22~ (1.00)$ \\
KLOE  & $503.1\pm 7.1~ (0.97)$  & $ 24.56 \pm 0.26 \pm 0.22~ (0.92)$   \\
CMD2  & $506.6\pm 3.9~ (0.89)$  & $ 24.96 \pm 0.21 \pm 0.22~ (0.96)$   \\
SND   & $505.1\pm 6.7~ (0.94)$  & $ 24.82 \pm 0.30 \pm 0.22~ (0.91)$  \\
\noalign{\smallskip}\hline
\end{tabularx}
}
\end{table}
Table~\ref{tab:resultspipi} quotes the specific contributions of the various 
$\ee\to\pp$ cross section measurements to \amuhadLO. Also given are the corresponding 
CVC-based $\tau\to\pim\piz\nut$ branching fraction predictions. The largest (smallest)
discrepancy of $2.7\,\sigma$ ($1.2\,\sigma$) between prediction and direct measurement 
is exhibited by KLOE (BABAR). It is interesting to note that the four $\amuhadLO[\pp]$ 
determinations agree within errors (the overall $\chi^2$ of 
their average amounts to 3.2 for 3 degrees of freedom), whereas significant discrepancies 
are observed in the corresponding spectral functions~\cite{g209}. The combined contribution, 
computed from local averages of the spectral function data, is subjected to local 
error rescaling in case of incompatibilities.

The contributions of the $J/\psi$ and $\psi(2S)$ resonances are obtained by numerically 
integrating the corresponding undressed Breit-Wigner lineshapes. 
The errors in the integrals are dominated by the knowledge of the corresponding
bare electronic width $\Gamma_{R\to ee}^0$.

\begin{figure}[t]
\begin{center}
\includegraphics[width=\figsize]{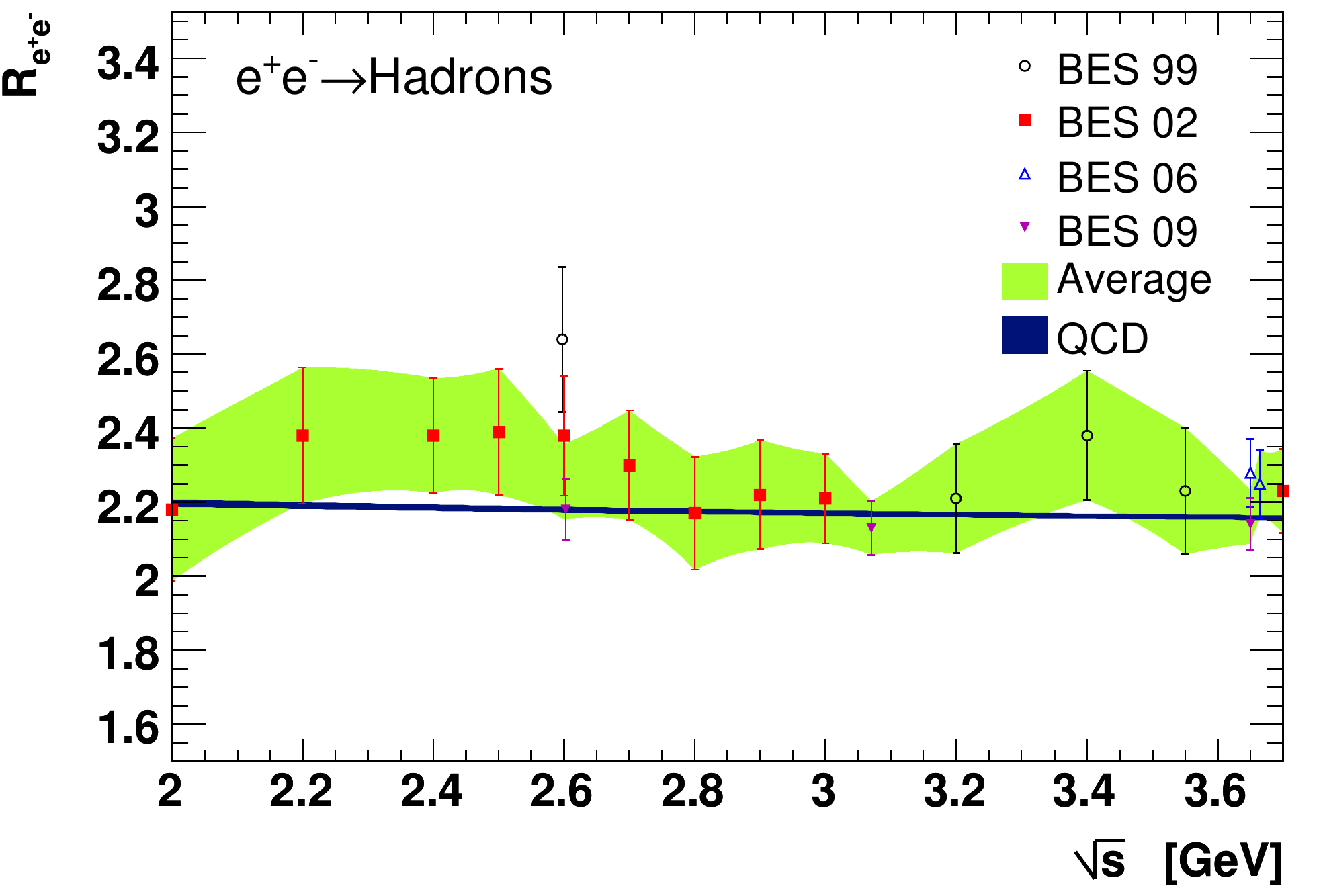}
\end{center}
\vspace{-0.4cm}
\caption[.]{ 
            Inclusive hadronic cross section ratio versus centre-of-mass energy in the 
            continuum region below the $D\Dbar$ threshold. Shown are bare
            BES data points~\cite{besR}, with statistical and systematic errors
            added in quadrature, the data average (shaded band), and the prediction 
            from massive perturbative QCD (solid line---see text). 
}
\label{fig:R}
\end{figure}
Sufficiently far from the quark thresholds four-loop~\cite{chetkuehn} perturbative 
QCD, including ${\cal O}(\as^2)$ quark mass corrections~\cite{kuhnmass}, is used 
to compute the inclusive hadronic cross section versus $\sqrt{s}$. Non-perturbative contributions at 
$1.8\,\gev$ were determined from data~\cite{dh98} and found to be 
small. The error in the perturbative prediction accounts for the
uncertainty in $\as$  ($\asZ=0.1193\pm0.0028$ from the fit to the $Z$ 
hadronic width~\cite{gfitter} is used), the truncation of the perturbative 
series (assigning the full four-loop contribution as systematic error), the full 
difference between fixed-order perturbation theory (FOPT) and, so-called, 
contour-improved perturbation theory (CIPT)~\cite{ledibpich}, as well as quark 
mass uncertainties (the values and errors from Ref.~\cite{pdg10} are used).
The former three errors are taken to be fully correlated between the 
various energy regions where perturbative QCD is used, whereas the (smaller) 
quark-mass uncertainties are taken to be uncorrelated. Figure~\ref{fig:R} shows 
the comparison between BES data~\cite{besR} and the QCD prediction below the $D\Dbar$ 
threshold between 2 and $3.7\,\gev$. Agreement within errors is found. Also for the 
transition region of $1.75$--$2.0\,\gev$, between the sum of exclusive measurements 
and QCD, excellent agreement between data and theory is found~\cite{dhmz2010}.

A full compilation of all contributions to \amuhadLO is given in Table~II of 
Ref.~\cite{dhmz2010}. 

\vspace{0.3cm}
\paragraph*{\bf\em Muon magnetic anomaly.}

Adding all lowest-order hadronic contributions together yields the estimate
(this and all following numbers in this and the next paragraph are in units 
of $10^{-10}$)~\cite{dhmz2010}
\beq
\label{eq:amuhadlo}
   \amuhadLO = 692.3 \pm 1.4 \pm 3.1 \pm 2.4 \pm 0.2 \pm 0.3\,,
\eeq
where the first error is statistical, the second channel-specific systematic, 
the third common systematic, correlated between at least two exclusive channels, 
and the fourth and fifth errors stand for the narrow resonance and QCD uncertainties, 
respectively. The total error of 4.2 is dominated by experimental systematic 
uncertainties. The new result is $-3.2\cdot 10^{-10}$ below the previous one~\cite{g209}. 
This shift is composed of
$-0.7$ from the inclusion of the new, large photon angle data from KLOE, 
$+0.4$ from the use of preliminary BABAR data in the $\ee\to\pp 2\piz$ mode, 
$-2.4$ from the new high-multiplicity exclusive channels, the re-estimate of 
the unknown channels, and the new resonance treatment, $-0.5$ from mainly the 
four-loop term in the QCD prediction of the hadronic cross section that 
contributes with a negative sign, as well as smaller other differences. 
The total error on \amuhadLO is slightly larger than that of Ref.~\cite{g209}
owing to a more conservative evaluation of the inter-channel correlations.

\begin{figure}[t]
\includegraphics[width=\columnwidth]{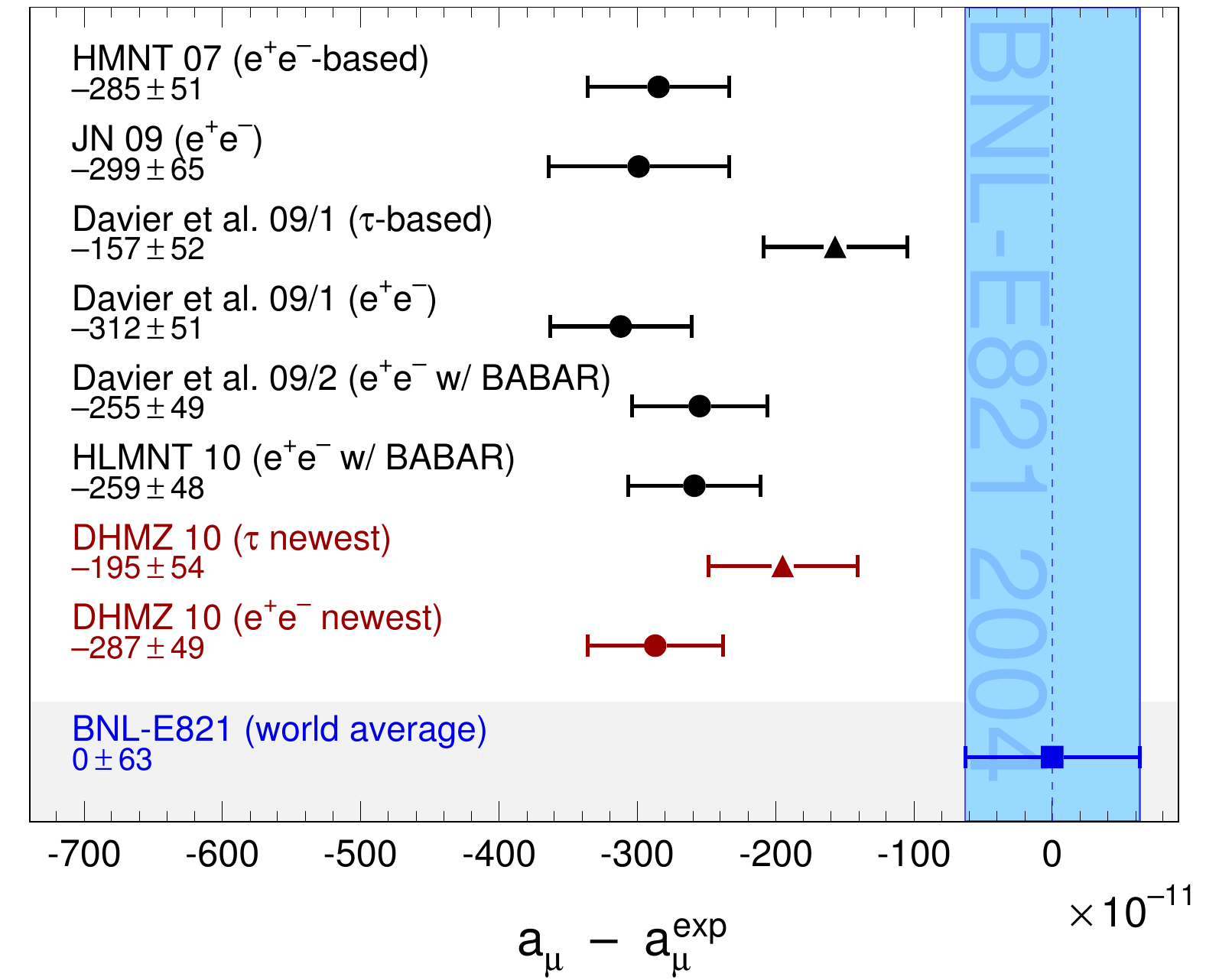}
\vspace{-0.4cm}
\caption{ 
        Compilation of recent results for $\amuSM$ (in units of $10^{-11}$),
        subtracted by the central value of the experimental average~(\ref{eq:amu_exp_num}).
        The shaded vertical band indicates the experimental error. 
        The SM predictions are taken from: 
        DHMZ 10~\cite{dhmz2010},
        HLMNT (unpublished)~\cite{hlmnt} (\ee based, including BABAR and KLOE 2010 \pp data),
        Davier \ea 09/1~\cite{eetaunew} (\Tau-based), 
        Davier \ea 09/1~\cite{eetaunew} (\ee-based, not including BABAR \pp data),
        Davier \ea 09/2~\cite{g209} (\ee-based including BABAR \pp data),
        HMNT 07~\cite{hmnt} and JN 09~\cite{jeger} (not including BABAR \pp data).}
\label{fig:amures}
\end{figure}
Adding to the result~(\ref{eq:amuhadlo}) the contributions from higher order 
hadronic loops, $-9.79 \pm 0.09$~\cite{hmnt}, computed using a similar dispersion 
relation approach, hadronic light-by-light scattering (LBLS), $10.5\pm 2.6$~\cite{prades09},  
estimated from theoretical model calculations (\cf remark in Footnote~\ref{ftn:lbls}), 
as well as QED~(\ref{eq:amu_qed_num}), and electroweak effects~(\ref{eq:amu_ew_num}), 
one obtains the full SM prediction
\beq
\label{eq:amusm}
  \amuSM = 11\,659\,180.2 \pm 4.2 \pm 2.6 \pm 0.2~(4.9_{\rm tot})\,,
\eeq
where the errors have been split into lowest and higher order hadronic, and 
other contributions, respectively. The result~(\ref{eq:amusm}) deviates from the 
experimental average~(\ref{eq:amu_exp_num}) by $28.7 \pm 8.0$ ($3.6\,\sigma$).\footnote
{\label{ftn:lbls}Using alternatively $11.6\pm 4.0$~\cite{nyffeler} for the light-by-light 
   scattering contribution, increases the error in the SM 
   prediction~(\ref{eq:amusm}) to $5.8$, and reduces the discrepancy with 
   experiment to $3.2\sigma$.
} 

A compilation of recent SM predictions for \amu compared with the experimental
result is given in Fig.~\ref{fig:amures}.

\vspace{0.3cm}
\paragraph*{\bf\em Update of \boldmath$\tau$-based $g-2$ result.}

Since the majority of the analysis in the \amu analysis also affects the 
$\tau$-based result from Ref.~\cite{eetaunew}, a reevaluation of the 
corresponding $\tau$-based hadronic contribution has been performed
in Ref.~\cite{dhmz2010}. In the $\tau$-based analysis~\cite{adh}, the 
\pp cross section is entirely replaced by the average, isospin-transformed, 
and isospin-breaking corrected $\tau\to\pim\piz\nut$ spectral function,\footnote
{
   Using published $\tau\to\pim\piz\nut$ spectral function data from 
   ALEPH~\cite{taualeph}, Belle~\cite{taubelle}, 
   CLEO~\cite{taucleo} and OPAL~\cite{tauopal}, and using the world 
   average branching fraction~\cite{pdg10} (2009 PDG edition).
}
while the four-pion cross sections, obtained from linear combinations of the 
$\taum\to\pim3\piz\nut$ and $\taum\to2\pim\pip\piz\nut$ spectral functions,
are only evaluated up to $1.5\,\gev$ with the $\tau$ data. Due to the lack of 
statistical precision, the spectrum is completed with the use of \ee data between 
1.5 and $1.8\,\gev$. All the other channels are taken from \ee data. The complete 
lowest-order $\tau$-based result reads~\cite{dhmz2010}
\beq
\label{eq:amuhadlotau}
   \amuhadLO[\tau] = 701.5 \pm 3.5 \pm 1.9 \pm 2.4 \pm 0.2 \pm 0.3\,,
\eeq
where the first error is $\tau$ experimental, the second estimates the uncertainty 
in the isospin-breaking corrections, the third is \ee experimental, and 
the fourth and fifth stand for the narrow resonance and QCD uncertainties, respectively. 
The $\tau$-based hadronic contribution differs by $9.1 \pm 5.0$ ($1.8\,\sigma$) from the 
\ee-based one, and the full $\tau$-based SM prediction $\amuSM[\tau]=11\,659\,189.4 \pm 5.4$
differs by $19.5 \pm 8.3$ ($2.4\,\sigma$) from the experimental average. This $\tau$-based 
result is also included in the compilation of Fig.~\ref{fig:amures}.

\vspace{0.3cm}
\paragraph*{\bf\em Running electromagnetic fine structure constant at \boldmath$M_Z^2$.}

The running electromagnetic fine structure constant, 
$\alpha(s)=\alpha(0)/(1-\Delta\alpha_{\rm lep}(s)-\Delta\alpha_{\rm had}(s))$,
at the scale of the $Z$ mass-squared, $s=M_Z^2$,
is an important ingredient of the SM fit to electroweak precision data at 
the $Z$ pole. Similar to \amu, the error on the $\alpha(M_Z)$ is dominated 
by hadronic vacuum polarisation. 

The sum of all the hadronic contributions gives for the \ee-based hadronic 
term in the running of \aZ
\beq
\label{eq:dahad}
   \dahadZ   = (274.2 \pm 1.0)\cdot 10^{-4}\,,
\eeq
which is, contrary to the evaluation of \amuhadLO, not dominated by the 
uncertainty in the low-energy data, but by contributions 
from all energy regions, where both experimental and theoretical errors 
are of similar magnitude. The corresponding $\tau$-based result reads 
$\dahadZ=(275.4 \pm 1.1)\cdot 10^{-4}$. As expected, the result~(\ref{eq:dahad}) 
is smaller than the most recent (unpublished) value from the HLMNT 
group~\cite{hlmnt} $\dahadZ=(275.2 \pm 1.5)\cdot 10^{-4}$. Owing to the use of 
perturbative QCD between 1.8 and $3.7\,\gev$, the precision in Eq.~(\ref{eq:dahad}) 
is  significantly improved compared to the HLMNT result, which relies 
on experimental data in that domain.\footnote
{
   HLMNT use perturbative QCD for the central value of the contribution 
   between 1.8 and $3.7\,\gev$, but assign the experimental errors from 
   the BES measurements to it. 
} 

Adding the three-loop leptonic contribution,
$\Delta\alpha_{\rm lep}(M_Z^2)=314.97686\cdot 10^{-4}$~\cite{steinhauser}, 
with negligible uncertainty, one finds
\beq
   \alpha^{-1}(M_Z^2) = 128.962 \pm 0.014\,.
\eeq

\begin{figure}[t]
\includegraphics[width=\columnwidth]{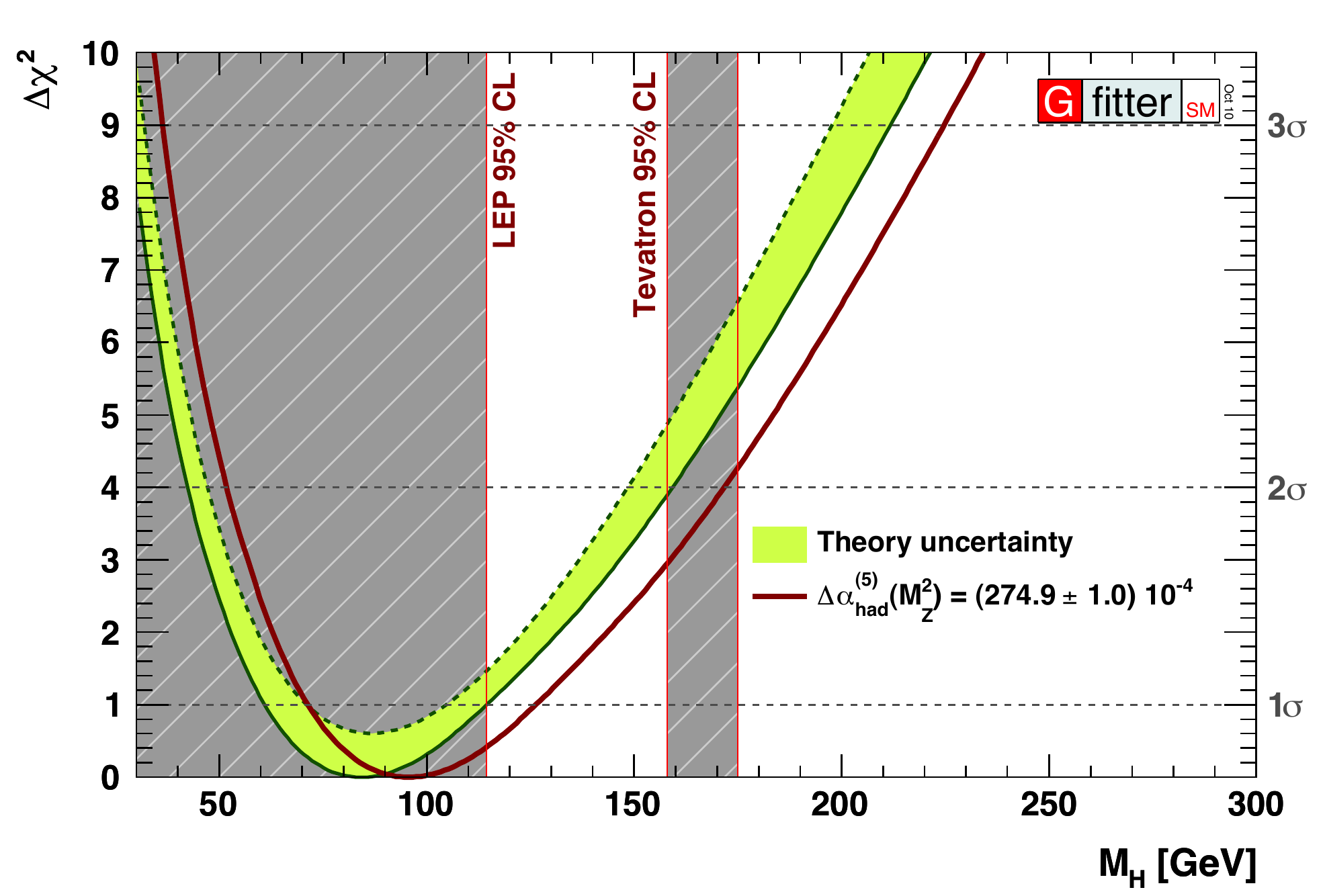}
\vspace{-0.4cm}
\caption{ 
         Standard Gfitter electroweak fit result~\cite{gfitter} (green 
         shaded band) and the result obtained for the new evaluation of 
         \dahadZ (red solid curve). Note that the legend displays the 
         corresponding five-quark contribution,
         \dahadZf, where the top term of $-0.72\cdot10^{-4}$ is excluded. 
         A shift of $+12\,\gev$ in the central value of the Higgs boson
         is observed.
         }
\label{fig:gfitter}
\end{figure}
The running electromagnetic coupling at $M_Z$ enters at various levels the 
global SM fit to electroweak precision data. It contributes to the radiator 
functions that modify the vector and axial-vector couplings in the partial 
$Z$ boson widths to fermions, and also to the SM prediction of the 
$W$ mass and the effective weak mixing angle. Overall, the fit exhibits
a $-39\%$ correlation between the Higgs mass ($M_H$) and $\dahadZ$~\cite{gfitter}, 
so that the decrease in the value~(\ref{eq:dahad}) and thus in the running
electromagnetic coupling strength with respect to earlier evaluations
leads to an increase in the best fit value for $M_H$.\footnote
{
   The correlation between $M_H$ and $\dahadZ$ reduces to $-17\%$ when 
   using the result~(\ref{eq:dahad}) in the global fit. 
} 
Figure~\ref{fig:gfitter} shows the standard
Gfitter result (green shaded band)~\cite{gfitter}, using as hadronic  
contribution $\dahadZ=(276.8\pm2.2)\cdot 10^{-4}$~\cite{hmnt}, together with 
the result obtained by using Eq.~(\ref{eq:dahad}) (red solid line).
The fitted Higgs mass shifts from previously $84^{\,+30}_{\,-23}\,\gev$ 
to $96^{\,+31}_{\,-24}\,\gev$. The larger error of the 
latter value, in spite of the improved accuracy in \dahadZ, is due to the 
logarithmic $M_H$ dependence of the fit observables. The new 95\% and 99\%
upper limits on $M_H$ are $170\,\gev$ and $201\,\gev$, respectively. 

\vspace{0.5cm}
\section{Conclusions}

Updated Standard Model predictions of the hadronic contributions to the 
muon anomalous magnetic moment and to the running electromagnetic coupling 
constant at $M_Z^2$ have been reported in Ref.~\cite{dhmz2010}. Mainly the 
reestimation of missing higher 
multiplicity channels, owing to new results from BABAR, causes a decrease 
of this contribution with respect to earlier evaluations, which---on one 
hand---amplifies the discrepancy of the muon $g-2$ measurement with its prediction 
to $3.6\,\sigma$ for \ee-based analysis, and to $2.4\,\sigma$ for the $\tau$-based
analysis, while---on the other hand---it relaxes the tension between the direct 
Higgs searches and the electroweak fit by $12\,\gev$ for the Higgs mass.

A thorough reestimation
of inter-channel correlations led to a slight increase in the final error
of the hadronic contribution to the muon $g-2$. A better precision is 
currently constricted by the discrepancy between KLOE and the other experiments,
in particular BABAR, in the dominant \pp mode. This discrepancy is corroborated
when comparing \ee and $\tau$ data in this mode, where agreement between 
BABAR and the $\tau$ data is observed. 

Support for the KLOE results must come from a cross-section measurement involving 
the ratio of pion-to-muon pairs. Moreover, new \pp precision data are soon expected 
from the upgraded VEPP-2000 storage ring at BINP-Novosibirsk, Russia, and the improved 
detectors CMD-3 and SND-2000. The future development of this field also relies 
on a more accurate muon $g-2$ measurement, and on progress in the evaluation of
the light-by-light scattering contribution. 

\begin{details}
I am grateful to the fruitful collaboration with my colleagues and friends 
Michel Davier, Bogdan Malaescu and Zhiqing Zhang. Martin 
Goebel from the Gfitter group is warmly thanked for performing the electroweak 
fit and producing Fig.~\ref{fig:gfitter} of these proceedings. I am indebted to 
George Lafferty and the helping hands at the University of Manchester who managed 
to organise a very interesting and pleasant workshop, and who so
thoughtfully provided umbrellas against boisterous rain storms.
\end{details}


\input refs